\documentclass[aps,showpacs,twocolumn,twoside,nofootinbib,amsmath,amssymb,superscriptaddress,prc]{revtex4-2}
\usepackage{multirow}
\usepackage{tikz}
\usepackage{epstopdf}
\usepackage[percent]{overpic}  
\usepackage{scrextend}  
\usepackage{xcolor,xspace}
\usepackage{todonotes}
\usepackage{hyperref}
\hypersetup{colorlinks=True,urlcolor=blue,linkcolor=blue,citecolor=blue,filecolor=black}

\usepackage{amsmath}
\usepackage{MnSymbol}
\usepackage{dcolumn,color,footnote,bm,braket}
\usepackage{url,longtable,tabularx}

\usepackage{fancybox}
\usetikzlibrary{matrix}

\begin{document}

\title{
High-$K$ isomers in a self-consistent mean-field approach with the Gogny force 
}

\author{L.~M.~Robledo}
\email{luis.robledo@uam.es}
\affiliation{Departamento de F\'\i sica Te\'orica and CIAFF, Universidad
Aut\'onoma de Madrid, E-28049 Madrid, Spain}

\affiliation{Center for Computational Simulation,
Universidad Polit\'ecnica de Madrid,
Campus de Montegancedo, Bohadilla del Monte, E-28660-Madrid, Spain}

\date{\today}

\begin{abstract}
High-$K$ isomeric states in even-even and odd-mass nuclei are described 
within a mean-field framework with full blocking and using the finite 
range Gogny force. Theoretical calculations of low energy spectra of 
several nuclei across the nuclear chart are compared with equal filling 
approximation results and experimental data. Despite the global 
character of the employed interactions, a good agreement between the 
different many-body methods and experimental data is found.
\end{abstract}

\maketitle


\section{Introduction}


The spectrum of atomic nuclei presents a rich variety of situations, 
ranging from collective states, where many nucleons participate 
coherently to the dynamic, to single particle excitations, where 
particles jump from  occupied orbits to unoccupied ones. The former are 
easier to identify as many of their characteristics are very robust as 
they are dictated by symmetries or the lack of them, and they 
traditionally lie down at low excitations energies. On the other hand, 
single particle excitations (one-particle one-hole excitations, 
two-particle two-hole, etc.) are far more numerous, but lie higher up 
in excitation energy and their electromagnetic decay is far weaker than 
in the collective case. Among the large variety of single particle 
excitations, high-$K$ isomers occupy a prominent place due to their 
large projection of angular momentum along the $z$ axis (the so-called 
$K$ quantum 
number)~\cite{Walker1999,Walker2001,Herzberg2006,Kondev2015,Dracoulis2016,Ackermann2017,Walker2020}. 
High-$K$ isomers require rather uncommon combinations of large $\Omega$ 
single-particle states close to the Fermi level %
\footnote{In the 
following, both $K$ and $\Omega$ represent the projection of angular 
momentum along $z$ axis. $K$ is for quasiparticle excitations while 
$\Omega$ is for single particle orbitals.}, %
and therefore they are 
only present in specific regions of the nuclear chart. In addition, 
their large $K$-value represent a strong hindrance in the 
electromagnetic transition strengths to the surrounding low-$K$ states, 
resulting in long lifetimes that facilitate their experimental 
characterization while providing an excellent target for theoretical 
studies. High-$K$ isomer physics can be of great importance in 
different scenarios like energy storage, therapeutic uses or the 
understanding of stellar nucleosynthesis 
\cite{Aprahamian2005,Garg2023}, just to mention a few. In the later 
case, the high temperature environment makes possible to populate 
high-$K$ isomeric states that act, thereby, as potential waiting points 
in the r-process mechanims - the so-called astromers \cite{Misch2021}. 
For instance, recent studies suggested that long-lived isomers could 
impact the kilonova light curve produced by the nucleosynthesis of 
heavy elements in neutron star 
mergers~\cite{Fujimoto2020b,Misch2021,Misch2021a}, calling for 
theoretical calculations of long-lived isomers in the neutron-rich 
region of the nuclear chart.

Mic-mac models based on a combination of a microscopic Woods-Saxon 
potential and a macroscopic deformation dependent liquid drop energy. 
Due to its simplicity, they are very popular in the description of many 
nuclear structure phenomena including high-$K$ isomers 
\cite{Nazarewicz1990,Jachimowicz2015}. Initial calculations of high-$K$ 
states were initially restricted to fixed deformation parameters (often 
assumed axially symmetric). The situation changed with the development 
of the multi-qp potential energy surface method \cite{Xu1998} (see Ref 
\cite{Jachimowicz2015} for a thorough application of the method in the 
super-heavy region). It allows a more flexible characterization of the 
deformation of isomeric states including triaxial effects. This 
flexibility is akin to the expected consequences of self-consistent 
blocking. In mic-mac models, pairing is often considered through a 
monopole pairing force treated at the Bardeen-Cooper-Schrieffer (BCS) 
level. In some calculations, the Lipkin-Nogami (LN) method is used to 
include dynamic correlations beyond mean field. The main difficulty of 
mic-mac models is the difficulty to obtain wave functions beyond those 
Wood-Saxon + BCS mean field ones. This represents a strong limitation in 
the calculation of the decay out of high-$K$ states. Also, the success 
of the method in describing experimental data relies on a carefully 
refitting of the parameters to specific and limited regions of the 
nuclear chart restricting its predicting power. 

The projected shell model (PSM) \cite{Hara1995} is a semi-microscopic 
theoretical tool often used to study high-$K$ isomers \cite{Sun2004}. 
It uses a combination of intrinsic multiquasiparticle excitations 
projected to good angular momentum to obtain sophisticated and highly 
correlated many body wave functions. Undoubtedly, this aspect 
represents and advantage in the description of the decay out of 
high-$K$ isomeric states. The multi-quasiparticle configurations are 
built on top of a common vacuum with fixed deformation and pairing gap 
and therefore they can only incorporate the effect of self-consistent 
blocking through the subsequent linear combination of excitations 
limiting the possibilities of rearrangement of deformation and pairing 
correlations of multi-quasiparticle excitations. This represents an 
important drawback of the method. The configuration space spans two 
major oscillator shells both for protons and neutron (not necessarily 
the same in both cases) and the pairing plus 
quadrupole (P+Q) Hamiltonian \cite{RS80} is used for the interaction. The model has proved 
to be very successful in describing high spin physics in various 
regions of the nuclear chart. Although the Hamiltonian is expected to 
embrace the two most relevant aspects of the nuclear residual 
interaction (quadrupole and pairing collectivities), it is constrained to deal with a fixed deformation 
parameter, limiting its applicability to situations where the different 
excitations of the system have similar deformations and pairing 
properties. This limitation prevents a more general characterization of 
the decay out of high-$K$ states.  Also, the parameters of the 
Hamiltonian are obtained from experimental data limiting its 
applicability in new regions of the nuclear chart. Recently \cite{Wu2017}
the PSM has been extended to handle up to ten-quasiparticle excitations
by using the pfaffian formula of \cite{Bertsch2012,PhysRevC.79.021302}.

Previous results concerning two-quasiparticles excitations in actinide and 
super-heavy nuclei obtained with Gogny D1S have already been discussed 
in~\cite{Delaroche2006}. Time reversal symmetry was preserved in most
of the applications and only in a few examples it was allowed to break.
In addition, four and higher number of quasiparticle excitations were not 
taken into account. Other topics discussed in the present paper (see below)
were not considered in that reference.

The purpose of this paper is to show how high-$K$ isomeric states can 
be successfully described at the mean-field level with the blocking 
method and density dependent finite range forces. Two and 
four-quasiparticles excitations are considered to illustrate the 
method, but the formalism can be applied to an arbitrary number of 
quasiparticles excitations. The mechanism responsible for the reduction 
of the excitation energy of the multiquasiparticle configurations as 
compared to the perturbative estimation is identified as the quenching 
of pairing correlations. The results obtained with the blocking formalism are 
compared with the equal filling approximation neglecting the time-odd 
terms of the functional. The excellent agreement between both 
methods suggests a minor role of time odd-field in the description of
excitation energies. The nuclei chosen to illustrate the method are all even-even 
nuclei and belong to the category of being well characterized experimentally. Most 
interesting applications considering astromers 
\cite{Misch2021,Misch2021a} will be deferred to future publications. As 
discussed below, the main advantages of the present proposal versus 
other approaches discussed above are the use of self-consistent 
blocking that allows for different deformation parameters and pairing 
properties for different excitations. Also, the universal character of 
the Gogny force and the good reproduction of experimental data in the 
considered nuclei give us confidence on the predictive power of the 
proposal. Last but not least, the wave functions obtained can be used 
in sophisticated calculations of the decay mechanism including symmetry 
restoration. The universal character of the Gogny force also allows for 
a consistent and same-quality description of the decay products 
(members of rotational band, triaxial configuration, other 
multi-quasiparticle excitations, etc) facilitating the interpretation 
of the decay out mechanism of high-$K$ isomers.

The paper is structured as follows. In Sec.~\ref{sec:method} we present 
the theoretical methods employed for the calculation of 
multi-quasiparticles excitations. In Sec.~\ref{sec:results} we present 
the results for high-$K$ isomers and low energy spectra for several 
nuclei across the nuclear chart. Finally, in Sec.~\ref{sec:conclusions} 
we summarize the main results and outlook future work.


\section{Theoretical methods \label{sec:method}}


The self-consistent description of high-$K$ isomers is based on the Hartree-Fock-Bogoliubov 
(HFB) approximation with blocking~\cite{RS80,BR86}. In the HFB method, the concept of quasiparticle
is introduced by defining quasiparticle creation and annihilation operators
\begin{equation}\label{eq:Wtransf}
\left(\begin{array}{c}
\beta\\
\beta^\dagger
\end{array}\right)=\left(\begin{array}{cc}
U^{+} & V^{+}\\
V^{T} & U^{T}
\end{array}\right)\left(\begin{array}{c}
c\\
c^\dagger
\end{array}\right)\equiv W^{+ }\left(\begin{array}{c}
c\\
c^\dagger
\end{array}\right) \,,
\end{equation}
as well as the corresponding HFB state $|\Phi\rangle$, vacuum to all 
the annihilation quasiparticle operators $\beta_{\mu}$, i.e. 
$\beta_{\mu}|\Phi\rangle=0$.  The label $\mu$ indexes the quasiparticle 
configurations and often contains quantum numbers like parity or 
projection of angular momentum along the intrinsic $z$ axis (the $K$ 
quantum number). As the HFB method does not preserve particle number 
symmetry, an important concept is ``number parity'' (NP) describing the 
parity (even or odd) of the number of particles in the different 
components making $|\Phi\rangle$ and its excitations. Number parity is 
a symmetry of the system and imposes a super-selection rule: wave 
functions with opposite NP values cannot be mixed together. An 
even-even nucleus has to be described by a HFB state with even number 
parity for both protons and neutrons. The quasiparticle operators have 
odd NP (they involve just creation and annihilation operators) and 
therefore $\beta^{+}_{\mu}|\Phi\rangle$ has opposite NP to that of 
$|\Phi\rangle$. Genuine excitations of a given system $|\Phi\rangle$ 
are then given by two-, four-, etc quasiparticles excitations whereas 
one-, three-, etc quasiparticles excitations correspond to an odd mass 
system if $|\Phi\rangle$ is an even number of particles wave function. 

The Bogoliubov amplitudes $U$ and $V$ are determined by using the 
variational principle on the HFB energy $E_{HFB} = \langle \Phi | 
\hat{H} |\Phi \rangle$ leading to the well know HFB 
equation~\cite{RS80,BR86}. The obtained quasiparticle energies $E_{\mu}$ are 
the ingredients entering the ``perturbative'' excitation energy 
$E_{\mu_{1}}+\cdots+E_{\mu_{M}}$ of a multi-quasiparticle (MQP) 
excitation $\beta^{+}_{\mu_{1}}\cdots \beta^{+}_{\mu_{M}} |\Phi 
\rangle$. This ``perturbative'' method is widely used in the 
literature, see Ref.~\cite{Delaroche2006} for an application in 
super-heavy nuclei with the Gogny force.

As the MQP excitations do not necesarily share the same properties
of $|\Phi\rangle $, it is necessary to use a self-consistent procedure where, for
each MQP excitation, the $U$ and $V$ amplitudes are determined by using the
variational principle on the MQP energy 
\begin{equation}
E_{\mu_{1},\ldots,\mu_{M}} = 
\langle \Phi | \beta_{\mu_{M}} \cdots \beta_{\mu_{1}}  \hat{H}
\beta^{+}_{\mu_{1}} \cdots \beta^{+}_{\mu_{M}} | \Phi \rangle \,.
\end{equation}
When used along with specific constrains on collective parameters this method 
provides also with potential energy surfaces (PES) for each specific MQP excitation
opening up the possibility to study the coupling with collective excitations
using the generator coordinate method (GCM) framework.

The best way to handle MQP excitations is by using the ``blocking'' procedure. 
In the standard HFB method, the matrix of contractions
\begin{equation}
\mathbb{R}=
\left(\begin{array}{cc}
\langle \Phi |\beta^\dagger_\mu  \beta_\nu |\Phi\rangle & \langle \Phi |\beta^\dagger_\mu  \beta^\dagger_\nu |\Phi\rangle \\
\langle \Phi |\beta_\mu  \beta_\nu |\Phi\rangle & \langle \Phi |\beta_\mu  \beta^\dagger_\nu |\Phi\rangle
\end{array}\right)
=
\left(\begin{array}{cc}
0 & 0 \\
0 & \mathbb{I}
\end{array}\right)	
\end{equation}
is connected to the generalized density matrix $\mathcal{R}$ 
through the $W$ transformation~from Eq.~\eqref{eq:Wtransf}:
\begin{equation}
\mathcal{R} =
\left(\begin{array}{cc}
\langle \Phi |c^\dagger_k  c_l |\Phi\rangle & \langle \Phi |c^\dagger_k  c^\dagger_l |\Phi\rangle \\
\langle \Phi | c_k  c_l |\Phi\rangle & \langle \Phi |c_k  c^\dagger_l |\Phi\rangle
\end{array}\right)
=
\left(\begin{array}{cc}
\rho & \kappa \\
-\kappa^* &  1-\rho^*
\end{array}\right)
=
W
\mathbb{R}
W^\dagger \,.
\end{equation}
In the MQP case one has to replace $|\Phi\rangle$ by $|\tilde{\Phi}\rangle=\beta^{+}_{\mu_{1}}\cdots \beta^{+}_{\mu_{M}} |\Phi \rangle${}
which is again a HFB wave function but vacuum of a different set of quasiparticle operators. Therefore, Wick's theorem
is also valid  but the contraction matrix is now given by a different expression. In the one-quasiparticle case one has
\begin{equation}
	\mathbb{R}_\mu=
\left(\begin{array}{cc}
{\mathbb{I}_\mu} & 0 \\
0 & \mathbb{I}-{\mathbb{I}_\mu}
\end{array}\right){}
\end{equation}
where the notation
\begin{equation}
	(\mathbb{I}_\mu)_{\sigma \rho} = 1  \quad \textrm{if} \quad \mu=\sigma=\rho \,; \quad \textrm{0 otherwise}\,,
\end{equation} 
has been introduced. The key point of the ``blocking'' method is the 
decomposition
\begin{equation}
	\mathbb{R}_{\mu} = \mathbb{S}_\mu \mathbb{R} \mathbb{S}^\dagger_\mu \,,
\end{equation}
where the ``swap'' matrix
\begin{equation}
\mathbb{S}_\mu=
\left(\begin{array}{cc}
\mathbb{I}-\mathbb{I}_\mu & \mathbb{I}_\mu \\
\mathbb{I}_\mu & \mathbb{I}-\mathbb{I}_\mu
\end{array}\right){}
\end{equation}
has been introduced. When this transformation is applied to the
right of a given $W$ Bogoliubov matrix
\begin{equation}
	W_{\mu} = W \mathbb{S}_{\mu} \,,
\end{equation}
the column $\mu$ of $U$ and $V^{*}$ gets swapped. The ``swap'' matrix allows
to write the density matrix contraction
\begin{equation}
\begin{split}
\mathcal{R}_{\mu} = & 
\left(\begin{array}{cc}
\langle \tilde{\Phi} |c^\dagger_k  c_l |\tilde{\Phi}\rangle & \langle \tilde{\Phi} |c^\dagger_k  c^\dagger_l |\tilde{\Phi}\rangle \\
\langle \tilde{\Phi} | c_k  c_l |\tilde{\Phi}\rangle & \langle \tilde{\Phi} |c_k  c^\dagger_l |\tilde{\Phi}\rangle
\end{array}\right) \\
= &
W
\mathbb{R}_{\mu}
W^\dagger	
=
W
\mathbb{S}_\mu \mathbb{R} \mathbb{S}^\dagger_\mu
W^\dagger	
=
W_{\mu}\mathbb{R}W^{+}_{\mu} \,,
\end{split}
\end{equation}
that thereby justifies keeping the same formalism as in the ``fully paired'' case
but swapping column $\mu$ of the $U$ and $V^{*}$ matrices. The generalization
to a MQP excitation is straightforward
\begin{equation}
\begin{split}
	\mathbb{R}_{\mu_{1},\ldots,\mu_{M}}= & 
\left(\begin{array}{cc}
	\sum_{k=1}^{M} \mathbb{I}_{\mu_{k}} & 0 \\
	0 & \mathbb{I}-\sum_{k=1}^{M} \mathbb{I}_{\mu_{k}}
\end{array}\right) \\ 
= & (\prod_{k=1}^{M} \mathbb{S}_{\mu_{k}}) \mathbb{R} (\prod_{k=M}^{1} \mathbb{S}^\dagger_{\mu_{k}}) \,,
\end{split}
\end{equation}
leading to the definition 
\begin{equation}
	W_{\mu_{1},\ldots,\mu_{M}} = W (\prod_{k=1}^{M} \mathbb{S}_{\mu_{k}}) \,.
\end{equation}
This expression is equivalent to the swapping of columns $\mu_{1},\ldots,\mu_{M}$ in the $U$
and $V^{*}$ matrices. Interestingly, this result is in agreement with recent
finding \cite{Kasuya2020} showing that the number parity of an HFB state is
given by the determinant of the associated $W$ Bogoliubov matrix. Noticing that $\det \mathbb{S}_{\mu_{k}}=-1$
it is clear that, as expected, the number parity of the multi-quasiparticle excitation is
the one of the initial state times $(-1)^{M}$.

In this work, the calculations are performed using the density 
dependent finite range Gogny interaction. The density dependence comes 
in the form of a non-integer power of the spatial density corresponding 
to the HFB state under consideration and introduces a dependence of the 
interaction on the state considered. In the present case, the HFB state 
is a MQP excitation built on a reference HFB wave function 
$|\Phi\rangle$ and the spatial density to use in the density dependent 
terms is given by 
\begin{equation}
\rho (\vec{R})_{\mu_{1},\ldots,\mu_{M}} = \langle \Phi | \beta_{\mu_{M}} \cdots \beta_{\mu_{1}}  \hat{\rho} (\vec{R})
\beta^{+}_{\mu_{1}}\cdots \beta^{+}_{\mu_{M}} |\Phi \rangle \,.
\end{equation}
This choice is consistent with the fact that the energy of the MQP excitation must be given,
at zero order, by the sum of the HFB energy of the reference state 
$E_{0}=\langle \Phi | \hat H | \Phi\rangle$ plus the sum  of quasiparticle 
energies $E_{\mu_{1}}+\cdots+E_{\mu_{M}}$ relative to $|\Phi\rangle$. In order
to check this property one possibility is to  evaluate $ \rho (\vec{R})_{\mu_{1},\ldots,\mu_{M}} $  using the quasiparticle representation of 
the one-body density operator 
$\hat{\rho} (\vec{R})=\sum_{i=1}^{A} \delta (\vec{R}-\vec{r}_{i})$
with respect to $|\Phi\rangle$ \cite{RS80}
\begin{equation}
\hat{\rho} (\vec{R}) = \langle \Phi | \hat{\rho} (\vec{R}) |\Phi \rangle + 
\sum_{\sigma \sigma'} \rho^{11}_{\sigma \sigma'} (\vec{R}) \beta^{+}_{\sigma} \beta_{\sigma'} + 
\frac{1}{2}(\rho^{20}+\rho^{02}) \,,
\end{equation}
where the $\rho^{20}$ ($\rho^{02}$) operator contain two creation (annihilation) quasiparticle
operators and therefore its mean value with respect to $|\Phi\rangle$ is zero.
Only the first two terms of the above expression 
contribute to $ \rho (\vec{R})_{\mu_{1},\ldots,\mu_{M}}$, being the final result
\begin{equation}
\rho (\vec{R})_{\mu_{1},\ldots,\mu_{M}}=\rho_{0} (\vec{R}) +
\sum_{\mu=\mu_{1}}^{\mu_{M}} \rho^{11}_{\mu,\mu} (\vec{R}) \,.
\end{equation} 
In the expression above, $\rho_{0}(\vec{R})$ corresponds to the density of the reference state $|\Phi\rangle$
\begin{equation}
\rho_{0} (\vec{R}) = \langle \Phi | \hat{\rho} (\vec{R}) |\Phi \rangle = 
\sum_{kl} \varphi^{*}_{k} (\vec{R}) \varphi_{l} (\vec{R}) \rho_{lk} \,,
\end{equation}
and $\rho^{11}_{\mu,\mu}$ is the diagonal
element of the $11$ matrix $O^{11}$ of the one-body operator with matrix 
elements $f_{ij} (\vec{R})$
\begin{equation}
\rho^{11}_{\mu,\mu} (\vec{R})=\sum_{ij} f_{ij} (\vec{R}) (U^*_{i\mu}
U_{j\mu} -V^*_{j\mu} V_{i\mu}) \,,
\end{equation}
with $f_{ij}(\vec{R})=\langle i|\delta (\vec{r}-
\vec{R})|j\rangle = \varphi^{*}_{i} (\vec{R})\varphi_{j} (\vec{R})$. 
Assuming that $\rho_{0} (\vec{R})$ is much larger than $\rho^{11}_{\mu,\mu}(\vec{R})$,
one can expand the $\rho^{\alpha}$ density dependent term as
\begin{equation}
	\rho (\vec{R})_{\mu_{1},\ldots,\mu_{M}}^{\alpha} = 
	\rho^{\alpha}_{0} (\vec{R}) + 
	\alpha \rho^{\alpha-1}_{0} (\vec{R}) 
	\sum_{\mu=\mu_{1}}^{\mu_{M}}\rho^{11}_{\mu,\mu} (\vec{R}) + \cdots{}
	\label{eq:expansion}
\end{equation}
In the derivation of the HFB equation for $|\Phi\rangle$ one has to include
a ``rearrangement'' term $\partial \Gamma_{kl}$ in the definition of the Hartree-Fock
(HF) Hamiltonian $h_{kl}$ to account for the variation of the density dependent
term when $|\Phi\rangle$ is varied~\cite{Robledo2019}. In order to be 
consistent, the same ``rearrangement'' term is included in the HF Hamiltonian
entering the definition of the matrix $H^{11}_{\mu \nu}$ leading, upon
diagonalization, to the definition of the quasiparticle energies $E_{\mu}$. It turns
out that the extra ``rearrangement'' term to be added to $H^{11}_{\mu \mu}$ is
$\alpha \langle \Phi | \rho^{\alpha-1}_{0} (\vec{R}) \rho^{11}_{\mu,\mu} (\vec{R}) | \Phi \rangle$,
which corresponds to the first order term in the expansion of Eq.~\eqref{eq:expansion}. 
See~\cite{Robledo2019} for a detailed derivation in the one-quasiparticle 
case.

\begin{figure}[htb!]
\begin{center}
\includegraphics[width=0.8\columnwidth]{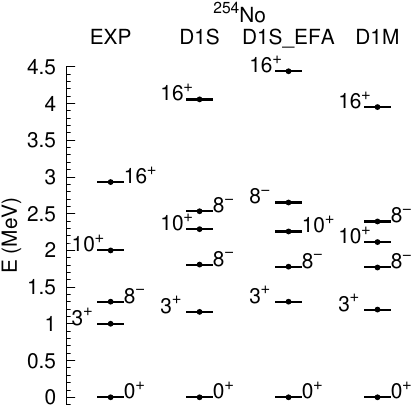}
\caption{Comparison of experimental \cite{Herzberg2006,Tandel2006,Clark201019,Garg2023} and theoretical results for two- and four-quasiparticles isomeric states 
in the nucleus $^{254}$No. The calculations have been carried out with the D1S and D1M parameterization
of the Gogny force and the full blocking procedure and EFA (see text for more details).} 
\label{fig:254No}
\end{center}
\end{figure}

\begin{figure}[htb!]
\begin{center}
\includegraphics[width=0.8\columnwidth]{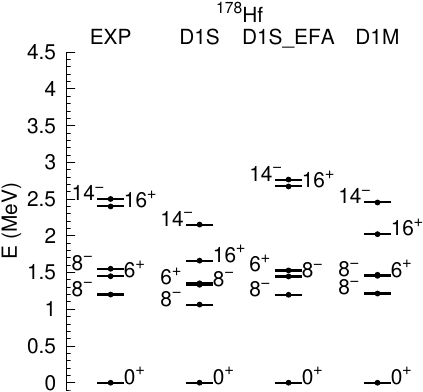}
\caption{Comparison of experimental \cite{Garg2023} (and references therein) and theoretical results for two- and four-quasiparticles isomeric states 
in the nucleus $^{178}$Hf. The calculations have been carried out with the D1S and D1M parameterization
of the Gogny force and the full blocking procedure and EFA (see text for more details).} 
\label{fig:178Hf}
\end{center}
\end{figure}

\subsection{The Gogny force}

As mentioned above, the interaction of choice is the effective 
density-dependent Gogny force, which is the sum of a finite-range
central potential, a sum of two Gaussians with different ranges, a zero-range 
two-body spin-orbit potential, a density-dependent term and Coulomb potential 
for protons~\cite{Robledo2019}. The parameters of the interaction are chosen as to reproduce 
basic nuclear matter properties and binding energies of finite nuclei.
There are essentially two parametrizations of the Gogny force traditionally employed in nuclear structure calculations (D1S~\cite{Berger1984} and D1M~\cite{Goriely2009}),
whose parameters depend on the targets used in the fitting protocol. Recently,
D1M*, a variant of D1M improving the symmetry energy properties, has been proposed~\cite{GonzalezBoquera2018}.
Other recent proposals include the D2 force~\cite{Chappert2015}, with its finite range density dependent term,
and the three-Gaussians variant D3G3~\cite{Batail2023}. In the following, we will focus
our attention in the more traditional D1S and D1M versions.


\subsection{Orthogonality}

As the ``blocking" procedure is variational, it is often the case that 
two calculations starting with different multi-quasiparticle 
excitations having the same $K$ and parity values end up in the same 
solution. Only in those cases where the starting MQP configurations 
differ in their deformation and pairing properties it is likely to find 
two differentiated solutions. The general solution to this problem is 
to introduce an orthogonality constraint in the self-consistent 
procedure. Fortunately, the number of possible configurations in 
high-$K$ isomeric states is very limited (often just one) reducing the 
chances to find orthogonality issues in the calculations. The lower the 
$K$ value of the excitation is the higher are the chances to hit this 
problem. As discussed below, the situation is specially critical for 
$K=0$ states. The study of low $K$ excitations is a very interesting 
issue that deserves further consideration.  

\subsection{Equal filling approximation}

The formalism presented above breaks time reversal invariance as the 
MQP excitation is not invariant under the action of the time reversal 
operator (converting the $K$ quantum number into $-K$). This 
characteristic translates to the density matrix and pairing field 
entering the HFB equation and therefore time-odd contributions have to 
be considered in the Hartree-Fock (HF) and pairing fields. This is not 
a problem for the Gogny force, as the time-odd fields come directly 
from the interaction itself, but it can represent a problem for density 
functional such as some Skyrme variants or the Barcelona Catania Paris 
Madrid (BCPM)~\cite{Baldo2023}, which are not considering those 
contributions. 

There is a formulation of the problem coined as the ``equal filling 
approximation''  (EFA) that preserves time reversal invariance and 
avoids time-odd fields characteristic of full blocking. This 
formulation was introduced heuristically many years ago to handle 
odd-mass nuclear systems and it was finally explained in terms of 
quantum statistical admixtures with specific probabilities in 
Ref.~\cite{PerezMartin2008}. The EFA is known to provide similar 
results as the ones obtained in the full blocking formulation 
\cite{Schunck2010,Giuliani2023} in the description of odd-$A$ systems, 
demonstrating the minor role played by time-odd fields. The 
modification of the spectrum with respect to the perturbative one is 
mostly due to the quenching of pairing correlations in the two cases. 
The formulation of the EFA in terms of quantum statistical admixtures 
allows to generalize the EFA concept to the present case of MQP 
excitations. This generalization was discussed in detail in 
Ref.~\cite{Giuliani2023}, and here we only recap the key concepts.

The density matrix for a multiquasiparticle excitation is given by  
\begin{equation}
\begin{split}\label{eq:roBmqp}
\rho_{kk'}^{(\mu_{B_{1}},\ldots,\mu_{B_{N}})} & =
\langle\Phi| \left( \prod_{\sigma=N}^{1}\beta_{\sigma} \right) c_{k'}^{\dagger}c_{k}
\left( \prod_{\sigma=1}^{N}\beta_{\sigma}^{\dagger} \right) |\Phi\rangle \\ 
& =\left(V^{*}V^{T}\right)_{kk'} + \sum_{\sigma} \left(U_{k'\sigma}^{*}U_{k\sigma}-V_{k'\sigma}V_{k\sigma}^{*}\right) \,,
\end{split}
\end{equation}
where $\sigma=\{ \mu_{B_{1}},\ldots,\mu_{B_{N}} \}$ for a 
$N$-quasiparticle excitation. The EFA expression for the 
multi-quasiparticle excitation density is obtained from 
Eq.~\eqref{eq:roBmqp} by multiplying the sum on the right most term by 
one half and extending the sum on the label $\sigma$ to include the 
time reverse quantum numbers of the set 
$\mu_{B_{1}},\ldots,\mu_{B_{N}}$. The same consideration applies 
straightforwardly to the EFA pairing tensor. Following the arguments of 
\cite{Giuliani2023}, it can be proved that the EFA density matrix and 
pairing tensors can be obtained  by introducing the statistical 
probabilities
\begin{equation}
p_{\sigma}=\left\{ \begin{array}{ccc}
		1 &  & \sigma \in {\mu_{B_{1}},\ldots,\mu_{B_{N}}};\,\,\textrm{{or}}\,\,\sigma \in {\overline{\mu}_{B_{1}},\ldots,\overline{\mu}_{B_{N}}}\\
0 &  & \textrm{otherwise}\end{array}\right. \,.\label{eq:P_EFAprobMQP}
\end{equation}

\subsection{Electromagnetic decay of high-$K$ isomers}

The electromagnetic decay of high-$K$ isomers involves multitude of 
different configurations including many multi-quasiparticles excitations 
as well as different members of collective rotational bands. Given the 
very different intrinsic properties of the high-$K$ isomers and the 
final states it is evident that the rotational formula commonly used to 
relate deformation parameters to transition probabilities 
\cite{Bohr1975} cannot be used here (see Ref \cite{Robledo2012} for an 
example of the failure of the rotational formula ). Therefore, it is 
mandatory to carry out the calculation using wave functions in the 
laboratory frame, i.e. using the intrinsic states projected to good 
angular momentum \cite{Sheikh2021}. This ambitious project is beyond the scope of the 
present work and will be addressed in the future.

\begin{figure}[htb!]
\begin{center}
\includegraphics[width=0.85\columnwidth]{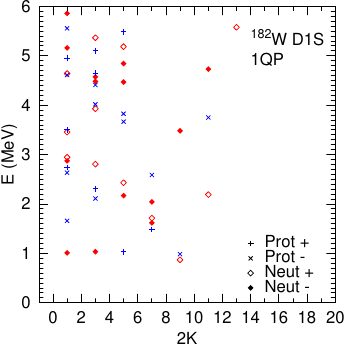}%
\caption{One-quasiparticle excitation energies (in MeV) in $^{182}$W for different values of the $K$ quantum number. Positive (negative) parity
	proton excitations are represented by the + ($\times$) symbol. For neutron excitations the symbols $\diamond$ and $\filleddiamond$ are used.}
\label{fig:182W1qp}
\end{center}
\end{figure}

\begin{figure}[htb!]
\begin{center}
\includegraphics[width=0.85\columnwidth]{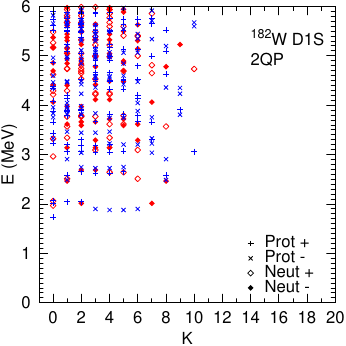}%
\caption{Two-quasiparticles excitation energies (in MeV) in $^{182}$W for different values of the combined $K$ quantum number. Positive (negative) parity
proton excitations are represented by the + ($\times$) symbol. For neutron excitations the symbols $\diamond$ and $\filleddiamond$ are used.} 
\label{fig:182W2qp}
\end{center}
\end{figure}

\section{Results and discussions\label{sec:results}}

There are many nuclei all over the nuclear chart with known high-$K$ 
isomeric states~\cite{Garg2023}. It is not the purpose of this paper to 
discuss in detail the properties of all of them, but rather to show 
that the  procedure discussed above works in a set of selected examples 
and provides a reasonable account of excitation energies of the high-$K$ 
states. One advantage in the calculation of high-$K$ states compared to 
low-$K$ states is that allows the neglection of orthogonality constraints.
This is because it is very unlikely to find two or more low-lying states 
with the same (high) $K$ values and similar mean-field properties (such as
deformation parameters, pairing, etc.). Therefore, in the following we 
restrict our study to high-$K$ isomeric states built on the ground state of
even-even nuclei, and postpone the discussion of odd-$A$ systems and 
fission isomers to a forthcoming publication.

\subsection{The $^{254}$No case}   

The heavy actinide $^{254}$No is a typical example where 
two-quasiparticles excitations of both protons and neutrons lead to 
high-$K$ isomeric states. In addition, four-quasiparticles excitations 
built on two-quasiparticles proton and neutron excitations lead to very 
high $K$ values for the associated states 
\cite{Herzberg2006,Tandel2006,Ackermann2017}. There are several known isomeric states 
in this nucleus being the $K=16^{+}$ four-quasiparticles excitation the 
best known example. This four-quasiparticles excitation with a half-life 
of 184$\,\,\mu s$ is made of  a two-quasiparticles excitation of protons in  
$\Omega=9/2$ and $\Omega=7/2$ orbitals together with a two-neutrons 
excitation from the same orbitals. In addition 
to this isomer there are others, like the $K=3^{+}$ (two-quasiparticles proton excitation) 
and two $K=8^-$ states (one corresponding to a two-protons excitation, and 
the other to a two-neutrons excitations). The $K=3^{+}$ state is obtained
from the excitation of protons in orbitals $\Omega=7/2$ and $\Omega=-1/2$ close
to the Fermi level and is the signature partner of a $K=4^{+}$ state not found
experimentally yet. For a detailed discussion of relevant single particle orbitals
in this nucleus obtained with different interactions the reader is referred
to Ref \cite{Dobaczewski2015}.
Additional two-neutrons 
quasiparticle states like the $K=10^{+}$ are observed 
\cite{Clark201019}. The experimental spectrum along with the results 
obtained with the Gogny D1S and D1M parametrizations are shown in Fig.~\ref{fig:254No}.
A careful analysis of the 
properties of the MQP excitations leads to the 
conclusion that pairing correlations are severly quenched in the 
isospin channel of the excitation. As a consequence, the four 
quasiparticle $K=16^+$ state shows almost no static pairing 
correlations. This general feature of MQP excitations call for a more 
developed and consistent treatment of dynamic pairing correlations 
\cite{Almehed2001}, by considering the effect of particle-number 
restoration before variation and fluctuations on the pairing-gap order 
parameter.

\begin{figure*}[htb!]
\begin{center}
\includegraphics[width=0.90\textwidth]{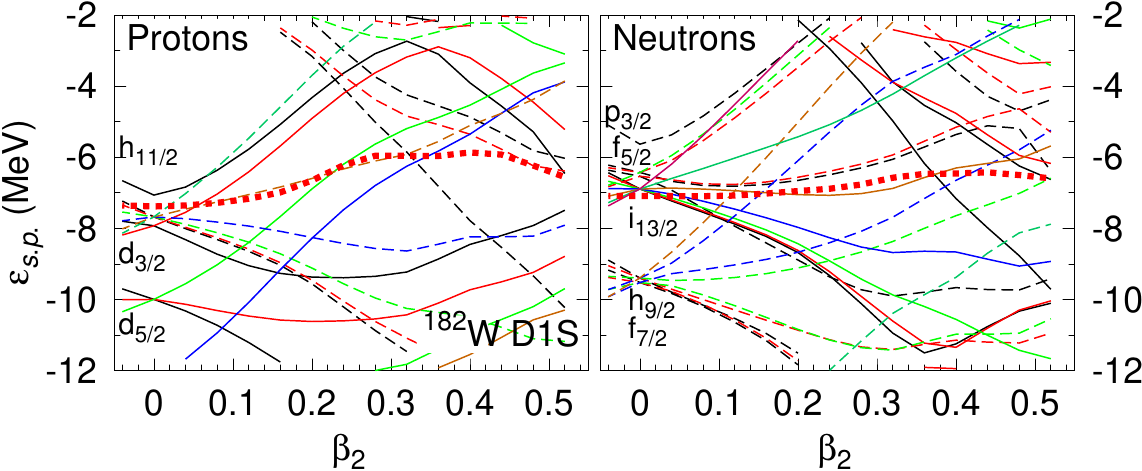}%
\caption{Single particle energies (in MeV) for protons and neutrons
in the isotope $^{182}$W are drawn as a function of the
$\beta_{2}$ deformation parameters. Full (dashed) lines correspond
to positive (negative) orbitals. The color code corresponds to the value
of the $\Omega$ quantum number of the single particle orbital: black, $\Omega=1/2$; 
red, $\Omega=3/2$; green, $\Omega=5/2$; blue, $\Omega=7/2$; etc. The $\Omega$ quantum number
of each orbital can also be inferred from the behavior of the single
particle energies near sphericity. The thick dotted line corresponds
to the Fermi level} 
\label{fig:182Wspe}
\end{center}
\end{figure*}

From these results, we can conclude that the impact of using D1S or D1M 
parametrization of the Gogny force is small for excitation energies, 
resulting in the same level of confidence in studying high-$K$ isomeric 
states.

It is also interesting to explore the results obtained with the EFA, 
discussed above. In Fig.~\ref{fig:254No} the results obtained with 
Gogny D1S employing both the blocking and EFA approximation are 
compared with the experimental data. The results of the EFA are 
qualitatively similar to the ones from full blocking indicating the 
little impact of time-odd fields in the excitation energies of the high 
$K$ states. As in the full blocking case discussed above, pairing 
correlations are severely quenched in the EFA. Additionally, one 
observes that the deformation parameters of both the blocking and EFA 
results are very close to each other. The good match between the full 
blocking and EFA results can represent a simplification in the 
theoretical calculations as the time-reversal preserving character of 
EFA simplifies the calculations. However, it has to be kept in mind 
that the EFA does not provide a wave function and therefore the 
calculation of transition probabilities requires further assumptions 
not present in the full-blocking case.

Finally, we conclude that the comparison between experimental data and 
theoretical calculations is satisfactory, particularly taking into 
account that the Gogny forces were not fitted specifically to 
reproduced single particle properties in the region of interest.

\subsection{The $^{178}$Hf case}

The high-$K$ isomers of the rare-earth nucleus $^{178}$Hf are also 
prototypical examples of high-$K$ states, mostly because the $16^{+}$ 
one has a lifetime of 31-yr, the longest among all the know high-$K$ 
isomeric states in even-even nuclei. Again, we observe several 
two-quasiparticles excitations of proton and neutron character as well 
as four-quasiparticles (two-protons and two-neutron) states. The 
lowest energy $8^{-}$ states is a two-neutrons excitation with 
particles being promoted to the $\Omega=9/2^{+}$ orbital (from the 
$i_{{13/2}}$), and the $\Omega=7/2^{-}$ orbital (from the spherical 
$h_{{9/2}}$). On the other hand, the next $8^{-}$ states is a 
two-protons excitation with particles being promoted to the 
$\Omega=9/2^{-}$ orbital (from the $h_{{11/2}}$) and the 
$\Omega=7/2^{+}$ orbital (from the $g_{{7/2}}$ spherical orbital). The 
combined excitation of the two-protons and two neutrons make the 
$K=16^{+}$ isomer. The assignments of the two $8^{-}$ states and the 
$16^{+}$ agree with the calculations of Ref.~\cite{Sun2004} using the 
projected shell model (PSM). Finally, the $6^{+}$ state is a 
two-protons excitation involving the $\Omega=5/2^{+}$ orbital (from the 
$d_{{5/2}}$) and the $\Omega=7/2^{+}$ orbital (from the $g_{{7/2}}$ 
spherical orbital). The  combined excitation of this proton $6^{+}$ 
excitation and the $8^{-}$ one discussed above is responsible for the 
$14^{-}$ state. The assignment of the $6^{+}$ state differs from the 
one of~\cite{Sun2004}, where it is claimed to be a two-neutrons 
excitation instead. The origin of the discrepancy between 
Ref.~\cite{Sun2004} and this work could be related to the many-body 
method used in the PSM calculations of~\cite{Sun2004}, which introduces 
correlations beyond mean-field not present in our approach. However, 
the PSM interaction employed in~\cite{Sun2004} is schematic and 
restricted to have a fixed value of quadrupole deformation parameters 
and pairing strengths for all the quasiparticle configurations. This 
limitation compares with the richness of the Gogny force in describing 
nuclear phenomena all over the nuclide chart~\cite{Robledo2019}. The 
assignment of deformed single particle orbitals to spherical orbits 
discussed in previous paragraphs can be obtained from 
Fig.~\ref{fig:182Wspe}, where the single particle spectrum as a 
function of quadrupole deformation parameter $\beta_{2}$ and for the 
nearby $^{182}$W isotope is displayed. Regarding the comparison of our 
results with experimental data in both D1S and D1M cases one can 
conclude that it is outstanding, giving credit to the claimed 
universalitiy of the Gogny interaction. Both parametrizations are able 
to reproduce the physics of high-$K$ isomers with the same set of 
parameters not only in the superheavies ($^{254}$No), but also in the 
rare-earth region. As in the nobelium case, pairing correlations are 
strongly suppressed and the quenching is responsible for the reordering 
of the spectrum as compared to the perturbative results obtained 
without selfconsistency. As mentioned before, the strong suppression of 
pairing suggests an appropriate treatment of dynamic pairing 
correlations.

As in the previous example, we have carried out calculations with Gogny 
D1S using the EFA. The results are compared with the ones obtained with 
full blocking and the experimental data  in Fig.~\ref{fig:178Hf}. As in 
the nobelium case, the results are very similar to those obtained with 
full blocking, indicating the minor role played by time-odd fields in 
the excitation energy of the isomers.

\subsection{The tungsten isotopic chain}

\begin{figure}[htb!]
\begin{center}
\includegraphics[width=0.90\columnwidth]{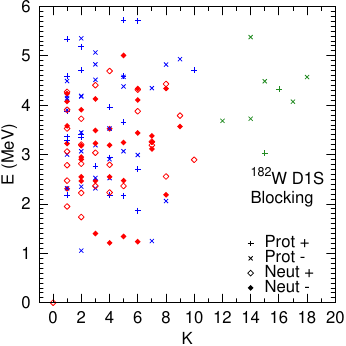}%
\caption{Self-consistent two-quasiparticles excitation energies (in MeV) for the 
$^{182}$W isotope. The assignment of each symbol (and color) is the 
same as in Fig.~\ref{fig:182W2qp}. Four-quasiparticles excitations 
composed of 2qp proton and 2qp excitations are also shown with green 
crosses (plus) symbols for negative (positive) parity states. 
The latter can be easily identified, as they are the 
only excitations with $K > 10$. }
\label{fig:182W}
\end{center}
\end{figure}

The region around $Z=72$ (Hf) and $N=106$ is known to have all the 
required characteristics  to show low lying high-$K$ isomeric states 
\cite{Walker2020}, namely the existence of high-$K$ single particle orbits
around the Fermi level. One of the species in the region with a large number 
of known isomers is tungsten ($Z=74$) and therefore its isotopes 
represent good candidates for high-$K$ isomer studies~\cite{Cullen1999}. 
As in previous examples, we restrict to even-even nuclei only and 
postpone the study of odd-$A$ isotopes to a forthcoming study. Additionally,
tungsten isotopes have been thoroughly studied with the PSM
approach~\cite{Wu2017,Jiao2012}.

\begin{figure*}[htb!]
\begin{center}
\includegraphics[width=0.31\textwidth]{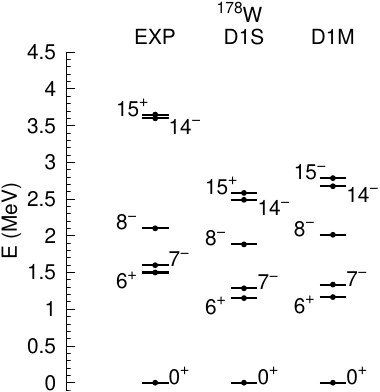}%
\includegraphics[width=0.31\textwidth]{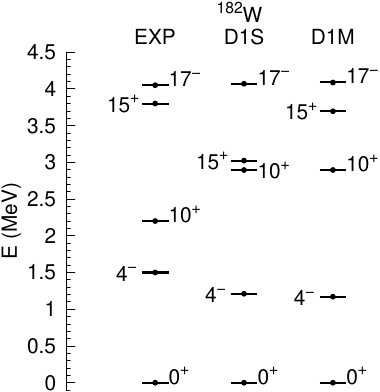}%
\includegraphics[width=0.31\textwidth]{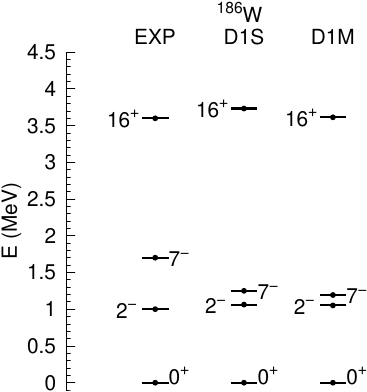}%
\caption{Comparison of theoretical results obtained with Gogny D1S and D1M,
and experimental data \cite{Regan1994,Shizuma1995,Purry1998,Baglin2003,Achterberg2009,Singh2010,Kondev2015,Garg2023} 
(and references therein), for relevant high-$K$ excitations known experimentally
in $^{178}$W, $^{182}$W, and  $^{186}$W.} 
\label{fig:Ww}
\end{center}
\end{figure*}

In Fig.~\ref{fig:182W1qp} we show, for the nucleus $^{182}$W, the 
lowest one-quasiparticle (1qp) excitation energies for both protons and 
neutrons labelled with both the $K$ and parity quantum numbers (no 
octupole correlations are present in this case). Five quasiparticles 
are located at around 1~MeV excitation energy, and two of them have 
specially large $K$ values: a $K^{\pi}=9/2^{+}$ neutron and a 
$K^{\pi}=9/2^{-}$ proton excitation. At slightly higher energies one 
observes three $K=7/2$ states, one for protons, two for neutrons. When 
combined together with the $K=9/2$ one-quasiparticle states they form 
high-$K$ low lying 2qp excitations. The complete perturbative spectrum 
of two-quasiparticle states is shown in Fig.~\ref{fig:182W2qp}. The 
perturbative spectrum is obtained by considering both positive and 
negative (i.e., the time reserved of the positive) $K$ values for each 
1qp state in order to obtain degenerated signature partners. In this 
plot, one observes the characteristic $\sim2$~MeV gap in excitation 
energy, which is the consequence of the rather strong pairing 
correlations both for protons (pairing gap $\Delta_{p}=0.71$ MeV) and 
neutrons ($\Delta_{n}=0.80$ MeV). Also, in the range of excitation 
energies below 6 MeV included in the plot, there are no 2qp excitations 
with $K$ larger than 10, indicating that higher $K$ excitations in the 
region must necessarily have a 4qp character. Many 2qp excitations are 
observed in the considered energy window, the number of them with a 
given $K$ value decreases with increasing $K$, being maximal for $K=0$ 
excitations.

In order to understand the characteristics of some of the discussed 
quasiparticle states it is convenient to look at the single particle 
energies, plotted as a function of the deformation parameter 
$\beta_{2}$ in Fig.~\ref{fig:182Wspe}. In separate panels the 
corresponding plots for both protons and neutrons are shown. Full 
(dashed) lines correspond to positive (negative) parity levels. The 
$\Omega$ values for each individual single particle state follow a 
color code (black for $\Omega=1/2$, red for $\Omega=3/2$, green for 
$\Omega=5/2$, etc.), but their values can also be inferred by looking at 
the splitting of spherical single particle levels when the quadrupole 
prolate deformation is switched on. At the typical deformation of the 
ground state of the tungsten isotopes considered, $\beta_{2}\approx 
0.25$, there is a positive parity state close to the Fermi level (the 
thick dashed line) with $\Omega^{\pi}=5/2^{+}$ (coming from the 
d$_{5/2}$ and with asymptotic  Nilsson quantum numbers [402]5/2), and a 
negative parity one with $\Omega^{\pi}=9/2^{-}$ (coming from the 
h$_{11/2}$ and with asymptotic Nilsson quantum numbers [514]9/2). In 
the neutron side, there is a $\Omega^{\pi}=9/2^{+}$ close to the Fermi 
level and coming from the i$_{13/2}$ orbital ([624]9/2), and two 
negative parity orbitals coming from the f$_{5/2}$ spherical orbital 
and $\Omega^{\pi}$ values of $1/2^{-}$ ([510]1/2) and $3/2^{-}$ 
([512]3/2). These single particle levels are responsible for the five 
1qp excitations with energies around 1~MeV discussed in
Fig.~\ref{fig:182W1qp}.

Selected 2qp and 4qp  excitations obtained after self-consistent full 
blocking are shown in Fig.~\ref{fig:182W} for the nucleus $^{182}$W. As 
the orthogonality constraint has not been imposed in the calculations 
many 2qp perturbative excitations converge, after the self-consistent 
procedure, to the lowest excited state with the same $K$ and parity 
values. This is the reason why one observes much less states than the 
ones shown in Fig.~\ref{fig:182W1qp}. This is particularly important 
for $K=0$ states, where only the ground state is obtained in the 
calculation if no orthogonality constraint is imposed. As the 
coexistence of multiple $K=0^+$ states is an interesting field of 
research by itself \cite{Meyer2006}, it will be the subject of a future 
study exploring the impact of the orthogonality constraint in MQP 
excitations. In the plot we show, along with the 2qp excitations, 
selected 4qp excitations with $K$ values larger than 10. In all the 
cases, the 4qp excitations correspond to a 2qp proton and a 2qp neutron 
multiquasiparticle excitation. The lowest 2qp excitation energies are 
slightly above 1~MeV, as compared to the perturbative 2qp excitation 
energies that are around 2~MeV. This reduction of almost one MeV 
indicates the importance of self-consistency in the determination of 
the excitation energies of the multiquasiparticle states. The origin of 
the excitation energy reduction can be traced back to the severe 
quenching of pairing correlation in the isospin channel of the 
excitation. As the ground state pairing gap for both types of nucleon 
is roughly 0.7~MeV for protons and 0.8~MeV for neutrons, its 
disappearance explains the almost 1~MeV reduction obtained in the 
blocking procedure.

Given that, in spite of the limitations imposed by not considering the 
orthogonality constraint, the number of 2qp excitations is huge, a 
comparison with experimental data is meaningless except for singular 
states, like the high-$K$ isomers under analysis. The singularity of 
those states resides in their large $K$ values, the very limited number 
of them and their long lifetimes. In order to compare with experimental 
data it is therefore necessary to proceed as in the previous cases and 
limit the comparison to those states that have been measured. In 
the theoretical side, one considers the lowest energy state with the 
same $K$ quantum number as the experimental one. Such a comparison is 
made in Fig.~\ref{fig:Ww} where the results of the calculations for 
$^{178}$W, $^{182}$W, and  $^{186}$W are shown.

In $^{178}$W, the $6^{+}$ is a two-neutrons excitation of orbitals with 
$\Omega=7/2$ and $\Omega=5/2$. The $7^{-}$ is also a two-neutrons 
excitation with two $\Omega=7/2$ orbitals of opposite parity involved.
On the other hand, the $8^{-}$ is a two-proton state made of a $\Omega=7/2${}
orbital and a $\Omega=9/2$ one. 
The $14^{-}$ and $15^{+}$ are four-quasiparticles excitations made of the 
two-protons excitation with $K=8^{-}$ and the $6^{+}$ and $7^{-}$ 
two-neutrons excitations, respectively. Our results agree with the
experimental data of Purry {\it et al } \cite{Purry1998} and the mic-mac
model resuls of Xu {\it et al} \cite{Xu1998}. In this nucleus there
are many more high-$K$ isomeric states known experimentally, including eight quasiparticle excitations made of
four-proton and four-neutron quasiparticle excitations \cite{Purry1998}. They will
be the subject of a more detailed study in the future. In $^{182}$W, both the $4^{-}$ and 
the $10^{+}$ are two-neutrons excitations, the first involving 
$\Omega=9/2$ and $\Omega=-1/2$ single particle orbitals and the other 
$\Omega=9/2$ and $\Omega=11/2$ ones. On the other hand, the $15^{+}$ 
and $17^{-}$ states are four-quasiparticles excitations, two-neutrons and 
two-protons. The two-protons excitation involves $\Omega=9/2$ and 
$\Omega=5/2$ orbitals providing $K=7$ and negative parity. The $17^{-}$ 
is the combination of the $K=7^{-}$ two-protons excitation and the
two-neutrons $K=10^{+}$ excitation, whereas the $15^{+}$ requires a 
$K=8^{-}$ two-neutrons excitation. Additionally, in $^{186}$W, the 
$K=7^{-}$ is the two-protons excitation observed in $^{182}$W. The 
$2^{-}$ is the partner of the $K=7^{-}$ with one of the orbitals 
reversed in time. Finally, the $16^{+}$ is made of the two-protons 
excitation with $K=7^{-}$ and a two-neutrons excitation with $K= 9^{-}$, 
consequence of the excitation of neutron single particle orbitals with 
$\Omega=7/2$ and $\Omega=11/2$.

The comparison with experimental data for the set of tungsten isotopes 
considered is very good, taking into account the global character of 
the family of Gogny forces considered in this paper. The limited number 
of parameters (14) is universal for all nuclei in the nuclear chart and 
all kind of phenomena. It is obvious then that such a global 
interaction cannot provide spectroscopic accuracy for the excitation 
energies of non-collective states at the level of other many-body 
methods and/or interactions. Spectroscopic quality can only be attained 
if specifically tailored-to-the-region interactions with tens or even 
hundred of parameters are used. One has also to take into account that 
the mean-field method being used in the present calculation  is not 
including beyond mean-field correlations like those stemming from 
symmetry restoration and/or particle-vibration coupling. Those 
correlations can amount to energy differences between states of 
hundreds of keV and, if they are not explicitly consider, it is not 
possible to extract a final conclusion regarding the quality of the 
interaction with respect to experimental data. Taking these 
considerations into account, one could consider satisfactory to obtain 
excitations energies differing by  hundred keV (or even one MeV) from 
experiment. This is because, in spite of these discrepancies, the 
predictions made by the present calculations can be a good guidance to 
experimental proposals and can also help to identify the origin of such 
excitations (deformation, pairing properties, etc.).

\subsection{Superheavy nuclei: the $^{270}$Ds case}

\begin{figure}[htb!]
\begin{center}
\includegraphics[width=0.90\columnwidth]{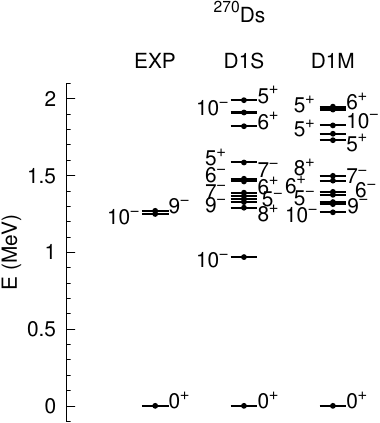}%
\caption{Comparison of high-$K$ isomeric states in $^{270}$Ds predicted by
Gogny D1S and D1M, and experimental
data~\cite{Hofmann2001,Ackermann2017}.} 
\label{fig:270Ds}
\end{center}
\end{figure}

In Hofmann {\it et al.}~\cite{Hofmann2001} the syntesis of the isotope  
of darmstadtium with $A=270$ and the identification of its decay products 
were reported. Along with the ground state, the decay of a high-$K$ 
isomer with a half-life in the milisecond range was identified. With 
this discovery, $^{270}$Ds is the heaviest nucleus where a high-$K$ 
isomeric state has been found~\cite{Ackermann2017}. The excitation energy of 
the isomer was estimated to be 1.13~MeV, in good agreement with results 
of HFB calculations reported in the same reference. The results of our 
calculation with both D1S and D1M parametrizations of Gogny are shown 
in Fig.~\ref{fig:270Ds}, along with experimental data. The experimental 
excitation energy of both the $9^{-}$ and $10^{-}$ states 
\cite{Ackermann2017} is well reproduced in the two cases and those 
states are assigned to a two-neutrons excitation. Along with the known 
experimental isomers, a bunch of other two-quasiparticles isomeric 
states with $K$ greater than 5 are predicted by the calculations and 
shown in the plot with the purpose to demonstrate that there are many 
more predicted isomeric states than the ones measured experimentally. 
For instance, the lowest $5^{-}$, $6^{-}$ and $7^{-}$ are two-neutrons 
excitations involving a $\Omega=11/2$ orbital. On the other hand, the 
lowest $6^{+}$ is a two-protons excitation. The predicted states 
indicate that $^{270}$Ds is a good candidate for experimental search of 
isomeric states. In the calculation, the quadrupole $\beta_{2}$ 
deformation parameter of the ground state and all isomeric states is 
essentially the same (up to one or two percent), with a value 
$\beta_{2}=0.245$. Reflection asymmetry is not present in the states 
displayed.

\section{Conclusions and Outlook\label{sec:conclusions}}

In this paper we present blocking calculations of two- and 
four-quasiparticle excitations leading to high-$K$ isomeric states in 
several relevant examples across the nuclear chart. The blocking 
procedure is presented in detail by using a novel approach involving 
``swap'' matrices. For the interaction, the Gogny force with the D1S 
and D1M parametrizations is used. The density prescription concerning 
density-dependent interactions is discussed, and it is shown that the 
``rearrangement effects'' in the definition of quasiparticle energies 
shall be taken into account. As typical cases in different regions of 
the nuclear chart we have carried out calculations in the super-heavy 
$^{254}$No, the rare-earth $^{178}$Hf, several tungsten isotopes and 
the super-heavy $^{270}$Ds. The agreement with experimental data is 
very good, particularly in the light of the parameter-free character of 
these calculations. The Gogny interaction is a global EDF adjusted to 
bulk nuclear properties, designed to provide a reasonable description 
of all kind of nuclear properties all over the nuclear chart. 
Additional calculations with the equal filling approximation show an 
almost perfect matching with the blocking ones. This result indicates 
that the effect of time-odd fields in multi-quasiparticle excitations 
is minor, and most of the reduction in the excitation energy (as 
compared to the sum of one-quasiparticle excitations) arises from the 
quenching of pairing correlations. Finally, a detailed study and 
comparison with experimental data is carried out in a series of 
tungsten isotopes. The good agreement with experimental data obtained 
for excitation energies give us confidence that effective forces like 
Gogny can be confidently used to describe high-$K$ excitations all over 
the nuclear chart. The results open the door to the inclusion of 
sophisticated beyond mean-field effects such as symmetry restoration, 
that will allow for a better and more systematic treatment of the 
electromagnetic decay out of such high-$K$ isomeric states. The 
formalism also allows for the  inclusion of fluctuation in collective 
degrees of freedom opening the possibility to study particle-vibration 
coupling from a microscopic perspective. The possibility to easily 
implement an orthogonality constraint also opens the door to a 
systematic study of $0^{+}$ excited states in even-even nuclei.

\begin{acknowledgments}
	The author thanks S. Giuliani for
	useful discussions. This work has been supported by the Spanish
	Agencia Estatal de Investigaci{\'o}n (AEI) of the Ministry of Science
	and Innovation (MCIN) under grant agreement No.~PID2021-127890NB-I00.
\end{acknowledgments}

%


\begin{thebibliography}{51}%
\makeatletter
\providecommand \@ifxundefined [1]{%
 \@ifx{#1\undefined}
}%
\providecommand \@ifnum [1]{%
 \ifnum #1\expandafter \@firstoftwo
 \else \expandafter \@secondoftwo
 \fi
}%
\providecommand \@ifx [1]{%
 \ifx #1\expandafter \@firstoftwo
 \else \expandafter \@secondoftwo
 \fi
}%
\providecommand \natexlab [1]{#1}%
\providecommand \enquote  [1]{``#1''}%
\providecommand \bibnamefont  [1]{#1}%
\providecommand \bibfnamefont [1]{#1}%
\providecommand \citenamefont [1]{#1}%
\providecommand \href@noop [0]{\@secondoftwo}%
\providecommand \href [0]{\begingroup \@sanitize@url \@href}%
\providecommand \@href[1]{\@@startlink{#1}\@@href}%
\providecommand \@@href[1]{\endgroup#1\@@endlink}%
\providecommand \@sanitize@url [0]{\catcode `\\12\catcode `\$12\catcode
  `\&12\catcode `\#12\catcode `\^12\catcode `\_12\catcode `\%12\relax}%
\providecommand \@@startlink[1]{}%
\providecommand \@@endlink[0]{}%
\providecommand \url  [0]{\begingroup\@sanitize@url \@url }%
\providecommand \@url [1]{\endgroup\@href {#1}{\urlprefix }}%
\providecommand \urlprefix  [0]{URL }%
\providecommand \Eprint [0]{\href }%
\providecommand \doibase [0]{https://doi.org/}%
\providecommand \selectlanguage [0]{\@gobble}%
\providecommand \bibinfo  [0]{\@secondoftwo}%
\providecommand \bibfield  [0]{\@secondoftwo}%
\providecommand \translation [1]{[#1]}%
\providecommand \BibitemOpen [0]{}%
\providecommand \bibitemStop [0]{}%
\providecommand \bibitemNoStop [0]{.\EOS\space}%
\providecommand \EOS [0]{\spacefactor3000\relax}%
\providecommand \BibitemShut  [1]{\csname bibitem#1\endcsname}%
\let\auto@bib@innerbib\@empty
\bibitem [{\citenamefont {Walker}\ and\ \citenamefont
  {Dracoulis}(1999)}]{Walker1999}%
  \BibitemOpen
  \bibfield  {author} {\bibinfo {author} {\bibfnamefont {P.}~\bibnamefont
  {Walker}}\ and\ \bibinfo {author} {\bibfnamefont {G.}~\bibnamefont
  {Dracoulis}},\ }\href {https://doi.org/10.1038/19911} {\bibfield  {journal}
  {\bibinfo  {journal} {Nature}\ }\textbf {\bibinfo {volume} {399}},\ \bibinfo
  {pages} {35} (\bibinfo {year} {1999})}\BibitemShut {NoStop}%
\bibitem [{\citenamefont {Walker}\ and\ \citenamefont
  {Dracoulis}(2001)}]{Walker2001}%
  \BibitemOpen
  \bibfield  {author} {\bibinfo {author} {\bibfnamefont {P.~M.}\ \bibnamefont
  {Walker}}\ and\ \bibinfo {author} {\bibfnamefont {G.~D.}\ \bibnamefont
  {Dracoulis}},\ }\href {https://doi.org/10.1023/A:1013915200556} {\bibfield
  {journal} {\bibinfo  {journal} {Hyperfine Interactions}\ }\textbf {\bibinfo
  {volume} {135}},\ \bibinfo {pages} {83} (\bibinfo {year} {2001})}\BibitemShut
  {NoStop}%
\bibitem [{\citenamefont {Herzberg}\ \emph {et~al.}(2006)\citenamefont
  {Herzberg}, \citenamefont {Greenlees}, \citenamefont {Butler}, \citenamefont
  {Jones}, \citenamefont {Venhart}, \citenamefont {Darby}, \citenamefont
  {Eeckhaudt}, \citenamefont {Eskola}, \citenamefont {Grahn}, \citenamefont
  {Gray-Jones}, \citenamefont {Hessberger}, \citenamefont {Jones},
  \citenamefont {Julin}, \citenamefont {Juutinen}, \citenamefont {Ketelhut},
  \citenamefont {Korten}, \citenamefont {Leino}, \citenamefont {Leppanen},
  \citenamefont {Moon}, \citenamefont {Nyman}, \citenamefont {Page},
  \citenamefont {Pakarinen}, \citenamefont {Pritchard}, \citenamefont
  {Rahkila}, \citenamefont {Saren}, \citenamefont {Scholey}, \citenamefont
  {Steer}, \citenamefont {Sun}, \citenamefont {Theisen},\ and\ \citenamefont
  {Uusitalo}}]{Herzberg2006}%
  \BibitemOpen
  \bibfield  {author} {\bibinfo {author} {\bibfnamefont {R.-D.}\ \bibnamefont
  {Herzberg}}, \bibinfo {author} {\bibfnamefont {P.~T.}\ \bibnamefont
  {Greenlees}}, \bibinfo {author} {\bibfnamefont {P.~A.}\ \bibnamefont
  {Butler}}, \bibinfo {author} {\bibfnamefont {G.~D.}\ \bibnamefont {Jones}},
  \bibinfo {author} {\bibfnamefont {M.}~\bibnamefont {Venhart}}, \bibinfo
  {author} {\bibfnamefont {I.~G.}\ \bibnamefont {Darby}}, \bibinfo {author}
  {\bibfnamefont {S.}~\bibnamefont {Eeckhaudt}}, \bibinfo {author}
  {\bibfnamefont {K.}~\bibnamefont {Eskola}}, \bibinfo {author} {\bibfnamefont
  {T.}~\bibnamefont {Grahn}}, \bibinfo {author} {\bibfnamefont
  {C.}~\bibnamefont {Gray-Jones}}, \bibinfo {author} {\bibfnamefont {F.~P.}\
  \bibnamefont {Hessberger}}, \bibinfo {author} {\bibfnamefont
  {P.}~\bibnamefont {Jones}}, \bibinfo {author} {\bibfnamefont
  {R.}~\bibnamefont {Julin}}, \bibinfo {author} {\bibfnamefont
  {S.}~\bibnamefont {Juutinen}}, \bibinfo {author} {\bibfnamefont
  {S.}~\bibnamefont {Ketelhut}}, \bibinfo {author} {\bibfnamefont
  {W.}~\bibnamefont {Korten}}, \bibinfo {author} {\bibfnamefont
  {M.}~\bibnamefont {Leino}}, \bibinfo {author} {\bibfnamefont {A.-P.}\
  \bibnamefont {Leppanen}}, \bibinfo {author} {\bibfnamefont {S.}~\bibnamefont
  {Moon}}, \bibinfo {author} {\bibfnamefont {M.}~\bibnamefont {Nyman}},
  \bibinfo {author} {\bibfnamefont {R.~D.}\ \bibnamefont {Page}}, \bibinfo
  {author} {\bibfnamefont {J.}~\bibnamefont {Pakarinen}}, \bibinfo {author}
  {\bibfnamefont {A.}~\bibnamefont {Pritchard}}, \bibinfo {author}
  {\bibfnamefont {P.}~\bibnamefont {Rahkila}}, \bibinfo {author} {\bibfnamefont
  {J.}~\bibnamefont {Saren}}, \bibinfo {author} {\bibfnamefont
  {C.}~\bibnamefont {Scholey}}, \bibinfo {author} {\bibfnamefont
  {A.}~\bibnamefont {Steer}}, \bibinfo {author} {\bibfnamefont
  {Y.}~\bibnamefont {Sun}}, \bibinfo {author} {\bibfnamefont {C.}~\bibnamefont
  {Theisen}},\ and\ \bibinfo {author} {\bibfnamefont {J.}~\bibnamefont
  {Uusitalo}},\ }\href {https://doi.org/10.1038/nature05069} {\bibfield
  {journal} {\bibinfo  {journal} {Nature}\ }\textbf {\bibinfo {volume} {442}},\
  \bibinfo {pages} {896} (\bibinfo {year} {2006})}\BibitemShut {NoStop}%
\bibitem [{\citenamefont {Kondev}\ \emph {et~al.}(2015)\citenamefont {Kondev},
  \citenamefont {Dracoulis},\ and\ \citenamefont {Kibédi}}]{Kondev2015}%
  \BibitemOpen
  \bibfield  {author} {\bibinfo {author} {\bibfnamefont {F.}~\bibnamefont
  {Kondev}}, \bibinfo {author} {\bibfnamefont {G.}~\bibnamefont {Dracoulis}},\
  and\ \bibinfo {author} {\bibfnamefont {T.}~\bibnamefont {Kibédi}},\ }\href
  {https://doi.org/https://doi.org/10.1016/j.adt.2015.01.001} {\bibfield
  {journal} {\bibinfo  {journal} {Atomic Data and Nuclear Data Tables}\
  }\textbf {\bibinfo {volume} {103-104}},\ \bibinfo {pages} {50} (\bibinfo
  {year} {2015})}\BibitemShut {NoStop}%
\bibitem [{\citenamefont {Dracoulis}\ \emph {et~al.}(2016)\citenamefont
  {Dracoulis}, \citenamefont {Walker},\ and\ \citenamefont
  {Kondev}}]{Dracoulis2016}%
  \BibitemOpen
  \bibfield  {author} {\bibinfo {author} {\bibfnamefont {G.~D.}\ \bibnamefont
  {Dracoulis}}, \bibinfo {author} {\bibfnamefont {P.~M.}\ \bibnamefont
  {Walker}},\ and\ \bibinfo {author} {\bibfnamefont {F.~G.}\ \bibnamefont
  {Kondev}},\ }\href {https://doi.org/10.1088/0034-4885/79/7/076301} {\bibfield
   {journal} {\bibinfo  {journal} {Reports on Progress in Physics}\ }\textbf
  {\bibinfo {volume} {79}},\ \bibinfo {pages} {076301} (\bibinfo {year}
  {2016})}\BibitemShut {NoStop}%
\bibitem [{\citenamefont {Ackermann}\ and\ \citenamefont
  {Theisen}(2017)}]{Ackermann2017}%
  \BibitemOpen
  \bibfield  {author} {\bibinfo {author} {\bibfnamefont {D.}~\bibnamefont
  {Ackermann}}\ and\ \bibinfo {author} {\bibfnamefont {C.}~\bibnamefont
  {Theisen}},\ }\href {https://doi.org/10.1088/1402-4896/aa7921} {\bibfield
  {journal} {\bibinfo  {journal} {Physica Scripta}\ }\textbf {\bibinfo {volume}
  {92}},\ \bibinfo {pages} {083002} (\bibinfo {year} {2017})}\BibitemShut
  {NoStop}%
\bibitem [{\citenamefont {Walker}\ and\ \citenamefont
  {Podolyák}(2020)}]{Walker2020}%
  \BibitemOpen
  \bibfield  {author} {\bibinfo {author} {\bibfnamefont {P.}~\bibnamefont
  {Walker}}\ and\ \bibinfo {author} {\bibfnamefont {Z.}~\bibnamefont
  {Podolyák}},\ }\href {https://doi.org/10.1088/1402-4896/ab635d} {\bibfield
  {journal} {\bibinfo  {journal} {Physica Scripta}\ }\textbf {\bibinfo {volume}
  {95}},\ \bibinfo {pages} {044004} (\bibinfo {year} {2020})}\BibitemShut
  {NoStop}%
\bibitem [{\citenamefont {Aprahamian}\ and\ \citenamefont
  {Sun}(2005)}]{Aprahamian2005}%
  \BibitemOpen
  \bibfield  {author} {\bibinfo {author} {\bibfnamefont {A.}~\bibnamefont
  {Aprahamian}}\ and\ \bibinfo {author} {\bibfnamefont {Y.}~\bibnamefont
  {Sun}},\ }\href {https://doi.org/10.1038/nphys150} {\bibfield  {journal}
  {\bibinfo  {journal} {Nature Physics}\ }\textbf {\bibinfo {volume} {1}},\
  \bibinfo {pages} {81} (\bibinfo {year} {2005})}\BibitemShut {NoStop}%
\bibitem [{\citenamefont {Garg}\ \emph {et~al.}(2023)\citenamefont {Garg},
  \citenamefont {Maheshwari}, \citenamefont {Singh}, \citenamefont {Sun},
  \citenamefont {Goel},\ and\ \citenamefont {Jain}}]{Garg2023}%
  \BibitemOpen
  \bibfield  {author} {\bibinfo {author} {\bibfnamefont {S.}~\bibnamefont
  {Garg}}, \bibinfo {author} {\bibfnamefont {B.}~\bibnamefont {Maheshwari}},
  \bibinfo {author} {\bibfnamefont {B.}~\bibnamefont {Singh}}, \bibinfo
  {author} {\bibfnamefont {Y.}~\bibnamefont {Sun}}, \bibinfo {author}
  {\bibfnamefont {A.}~\bibnamefont {Goel}},\ and\ \bibinfo {author}
  {\bibfnamefont {A.~K.}\ \bibnamefont {Jain}},\ }\href
  {https://doi.org/https://doi.org/10.1016/j.adt.2022.101546} {\bibfield
  {journal} {\bibinfo  {journal} {Atomic Data and Nuclear Data Tables}\
  }\textbf {\bibinfo {volume} {150}},\ \bibinfo {pages} {101546} (\bibinfo
  {year} {2023})}\BibitemShut {NoStop}%
\bibitem [{\citenamefont {Misch}\ \emph
  {et~al.}(2021{\natexlab{a}})\citenamefont {Misch}, \citenamefont {Sprouse},\
  and\ \citenamefont {Mumpower}}]{Misch2021}%
  \BibitemOpen
  \bibfield  {author} {\bibinfo {author} {\bibfnamefont {G.~W.}\ \bibnamefont
  {Misch}}, \bibinfo {author} {\bibfnamefont {T.~M.}\ \bibnamefont {Sprouse}},\
  and\ \bibinfo {author} {\bibfnamefont {M.~R.}\ \bibnamefont {Mumpower}},\
  }\href {https://doi.org/10.3847/2041-8213/abfb74} {\bibfield  {journal}
  {\bibinfo  {journal} {The Astrophysical Journal Letters}\ }\textbf {\bibinfo
  {volume} {913}},\ \bibinfo {pages} {L2} (\bibinfo {year}
  {2021}{\natexlab{a}})}\BibitemShut {NoStop}%
\bibitem [{\citenamefont {Fujimoto}\ and\ \citenamefont
  {Hashimoto}(2020)}]{Fujimoto2020b}%
  \BibitemOpen
  \bibfield  {author} {\bibinfo {author} {\bibfnamefont {S.-i.}\ \bibnamefont
  {Fujimoto}}\ and\ \bibinfo {author} {\bibfnamefont {M.-a.}\ \bibnamefont
  {Hashimoto}},\ }\href {https://doi.org/10.1093/mnrasl/slaa016} {\bibfield
  {journal} {\bibinfo  {journal} {Mon. Not. R. Astron. Soc. Lett.}\ }\textbf
  {\bibinfo {volume} {493}},\ \bibinfo {pages} {L103} (\bibinfo {year}
  {2020})},\ \Eprint {https://arxiv.org/abs/2001.10668} {arXiv:2001.10668}
  \BibitemShut {NoStop}%
\bibitem [{\citenamefont {Misch}\ \emph
  {et~al.}(2021{\natexlab{b}})\citenamefont {Misch}, \citenamefont {Sprouse},
  \citenamefont {Mumpower}, \citenamefont {Couture}, \citenamefont {Fryer},
  \citenamefont {Meyer},\ and\ \citenamefont {Sun}}]{Misch2021a}%
  \BibitemOpen
  \bibfield  {author} {\bibinfo {author} {\bibfnamefont {G.~W.}\ \bibnamefont
  {Misch}}, \bibinfo {author} {\bibfnamefont {T.~M.}\ \bibnamefont {Sprouse}},
  \bibinfo {author} {\bibfnamefont {M.~R.}\ \bibnamefont {Mumpower}}, \bibinfo
  {author} {\bibfnamefont {A.~J.}\ \bibnamefont {Couture}}, \bibinfo {author}
  {\bibfnamefont {C.~L.}\ \bibnamefont {Fryer}}, \bibinfo {author}
  {\bibfnamefont {B.~S.}\ \bibnamefont {Meyer}},\ and\ \bibinfo {author}
  {\bibfnamefont {Y.}~\bibnamefont {Sun}},\ }\bibfield  {journal} {\bibinfo
  {journal} {Symmetry}\ }\textbf {\bibinfo {volume} {13}},\ \href
  {https://doi.org/10.3390/sym13101831} {10.3390/sym13101831} (\bibinfo {year}
  {2021}{\natexlab{b}})\BibitemShut {NoStop}%
\bibitem [{\citenamefont {Nazarewicz}\ \emph {et~al.}(1990)\citenamefont
  {Nazarewicz}, \citenamefont {Riley},\ and\ \citenamefont
  {Garrett}}]{Nazarewicz1990}%
  \BibitemOpen
  \bibfield  {author} {\bibinfo {author} {\bibfnamefont {W.}~\bibnamefont
  {Nazarewicz}}, \bibinfo {author} {\bibfnamefont {M.}~\bibnamefont {Riley}},\
  and\ \bibinfo {author} {\bibfnamefont {J.}~\bibnamefont {Garrett}},\ }\href
  {https://doi.org/https://doi.org/10.1016/0375-9474(90)90004-6} {\bibfield
  {journal} {\bibinfo  {journal} {Nuclear Physics A}\ }\textbf {\bibinfo
  {volume} {512}},\ \bibinfo {pages} {61} (\bibinfo {year} {1990})}\BibitemShut
  {NoStop}%
\bibitem [{\citenamefont {Jachimowicz}\ \emph {et~al.}(2015)\citenamefont
  {Jachimowicz}, \citenamefont {Kowal},\ and\ \citenamefont
  {Skalski}}]{Jachimowicz2015}%
  \BibitemOpen
  \bibfield  {author} {\bibinfo {author} {\bibfnamefont {P.}~\bibnamefont
  {Jachimowicz}}, \bibinfo {author} {\bibfnamefont {M.}~\bibnamefont {Kowal}},\
  and\ \bibinfo {author} {\bibfnamefont {J.}~\bibnamefont {Skalski}},\ }\href
  {https://doi.org/10.1103/PhysRevC.92.044306} {\bibfield  {journal} {\bibinfo
  {journal} {Phys. Rev. C}\ }\textbf {\bibinfo {volume} {92}},\ \bibinfo
  {pages} {044306} (\bibinfo {year} {2015})}\BibitemShut {NoStop}%
\bibitem [{\citenamefont {Xu}\ \emph {et~al.}(1998)\citenamefont {Xu},
  \citenamefont {Walker}, \citenamefont {Sheikh},\ and\ \citenamefont
  {Wyss}}]{Xu1998}%
  \BibitemOpen
  \bibfield  {author} {\bibinfo {author} {\bibfnamefont {F.}~\bibnamefont
  {Xu}}, \bibinfo {author} {\bibfnamefont {P.}~\bibnamefont {Walker}}, \bibinfo
  {author} {\bibfnamefont {J.}~\bibnamefont {Sheikh}},\ and\ \bibinfo {author}
  {\bibfnamefont {R.}~\bibnamefont {Wyss}},\ }\href
  {https://doi.org/https://doi.org/10.1016/S0370-2693(98)00857-0} {\bibfield
  {journal} {\bibinfo  {journal} {Physics Letters B}\ }\textbf {\bibinfo
  {volume} {435}},\ \bibinfo {pages} {257} (\bibinfo {year}
  {1998})}\BibitemShut {NoStop}%
\bibitem [{\citenamefont {Hara}\ and\ \citenamefont {Sun}(1995)}]{Hara1995}%
  \BibitemOpen
  \bibfield  {author} {\bibinfo {author} {\bibfnamefont {K.}~\bibnamefont
  {Hara}}\ and\ \bibinfo {author} {\bibfnamefont {Y.}~\bibnamefont {Sun}},\
  }\href {https://doi.org/10.1142/S0218301395000250} {\bibfield  {journal}
  {\bibinfo  {journal} {International Journal of Modern Physics E}\ }\textbf
  {\bibinfo {volume} {04}},\ \bibinfo {pages} {637} (\bibinfo {year}
  {1995})}\BibitemShut {NoStop}%
\bibitem [{\citenamefont {Sun}\ \emph {et~al.}(2004)\citenamefont {Sun},
  \citenamefont {Zhou}, \citenamefont {Long}, \citenamefont {Zhao},\ and\
  \citenamefont {Walker}}]{Sun2004}%
  \BibitemOpen
  \bibfield  {author} {\bibinfo {author} {\bibfnamefont {Y.}~\bibnamefont
  {Sun}}, \bibinfo {author} {\bibfnamefont {X.-R.}\ \bibnamefont {Zhou}},
  \bibinfo {author} {\bibfnamefont {G.-L.}\ \bibnamefont {Long}}, \bibinfo
  {author} {\bibfnamefont {E.-G.}\ \bibnamefont {Zhao}},\ and\ \bibinfo
  {author} {\bibfnamefont {P.~M.}\ \bibnamefont {Walker}},\ }\href
  {https://doi.org/https://doi.org/10.1016/j.physletb.2004.03.066} {\bibfield
  {journal} {\bibinfo  {journal} {Physics Letters B}\ }\textbf {\bibinfo
  {volume} {589}},\ \bibinfo {pages} {83} (\bibinfo {year} {2004})}\BibitemShut
  {NoStop}%
\bibitem [{\citenamefont {Ring}\ and\ \citenamefont {Schuck}(1980)}]{RS80}%
  \BibitemOpen
  \bibfield  {author} {\bibinfo {author} {\bibfnamefont {P.}~\bibnamefont
  {Ring}}\ and\ \bibinfo {author} {\bibfnamefont {P.}~\bibnamefont {Schuck}},\
  }\href@noop {} {\emph {\bibinfo {title} {The nuclear many body problem}}}\
  (\bibinfo  {publisher} {Springer, Berlin},\ \bibinfo {year}
  {1980})\BibitemShut {NoStop}%
\bibitem [{\citenamefont {Wu}\ \emph {et~al.}(2017)\citenamefont {Wu},
  \citenamefont {Ghorui}, \citenamefont {Wang}, \citenamefont {Sun},
  \citenamefont {Guidry},\ and\ \citenamefont {Walker}}]{Wu2017}%
  \BibitemOpen
  \bibfield  {author} {\bibinfo {author} {\bibfnamefont {X.-Y.}\ \bibnamefont
  {Wu}}, \bibinfo {author} {\bibfnamefont {S.~K.}\ \bibnamefont {Ghorui}},
  \bibinfo {author} {\bibfnamefont {L.-J.}\ \bibnamefont {Wang}}, \bibinfo
  {author} {\bibfnamefont {Y.}~\bibnamefont {Sun}}, \bibinfo {author}
  {\bibfnamefont {M.}~\bibnamefont {Guidry}},\ and\ \bibinfo {author}
  {\bibfnamefont {P.~M.}\ \bibnamefont {Walker}},\ }\href
  {https://doi.org/10.1103/PhysRevC.95.064314} {\bibfield  {journal} {\bibinfo
  {journal} {Phys. Rev. C}\ }\textbf {\bibinfo {volume} {95}},\ \bibinfo
  {pages} {064314} (\bibinfo {year} {2017})}\BibitemShut {NoStop}%
\bibitem [{\citenamefont {Bertsch}\ and\ \citenamefont
  {Robledo}(2012)}]{Bertsch2012}%
  \BibitemOpen
  \bibfield  {author} {\bibinfo {author} {\bibfnamefont {G.~F.}\ \bibnamefont
  {Bertsch}}\ and\ \bibinfo {author} {\bibfnamefont {L.~M.}\ \bibnamefont
  {Robledo}},\ }\href {https://doi.org/10.1103/PhysRevLett.108.042505}
  {\bibfield  {journal} {\bibinfo  {journal} {Phys. Rev. Lett.}\ }\textbf
  {\bibinfo {volume} {108}},\ \bibinfo {pages} {042505} (\bibinfo {year}
  {2012})}\BibitemShut {NoStop}%
\bibitem [{\citenamefont {Robledo}(2009)}]{PhysRevC.79.021302}%
  \BibitemOpen
  \bibfield  {author} {\bibinfo {author} {\bibfnamefont {L.~M.}\ \bibnamefont
  {Robledo}},\ }\href {https://doi.org/10.1103/PhysRevC.79.021302} {\bibfield
  {journal} {\bibinfo  {journal} {Phys. Rev. C}\ }\textbf {\bibinfo {volume}
  {79}},\ \bibinfo {pages} {021302} (\bibinfo {year} {2009})}\BibitemShut
  {NoStop}%
\bibitem [{\citenamefont {Delaroche}\ \emph {et~al.}(2006)\citenamefont
  {Delaroche}, \citenamefont {Girod}, \citenamefont {Goutte},\ and\
  \citenamefont {Libert}}]{Delaroche2006}%
  \BibitemOpen
  \bibfield  {author} {\bibinfo {author} {\bibfnamefont {J.-P.}\ \bibnamefont
  {Delaroche}}, \bibinfo {author} {\bibfnamefont {M.}~\bibnamefont {Girod}},
  \bibinfo {author} {\bibfnamefont {H.}~\bibnamefont {Goutte}},\ and\ \bibinfo
  {author} {\bibfnamefont {J.}~\bibnamefont {Libert}},\ }\href
  {https://doi.org/https://doi.org/10.1016/j.nuclphysa.2006.03.004} {\bibfield
  {journal} {\bibinfo  {journal} {Nuclear Physics A}\ }\textbf {\bibinfo
  {volume} {771}},\ \bibinfo {pages} {103} (\bibinfo {year}
  {2006})}\BibitemShut {NoStop}%
\bibitem [{\citenamefont {Blaizot}\ and\ \citenamefont {Ripka}(1986)}]{BR86}%
  \BibitemOpen
  \bibfield  {author} {\bibinfo {author} {\bibfnamefont {J.-P.}\ \bibnamefont
  {Blaizot}}\ and\ \bibinfo {author} {\bibfnamefont {G.}~\bibnamefont
  {Ripka}},\ }\href@noop {} {\emph {\bibinfo {title} {Quantum theory of finite
  systems}}}\ (\bibinfo  {publisher} {The MIT press},\ \bibinfo {year}
  {1986})\BibitemShut {NoStop}%
\bibitem [{\citenamefont {Kasuya}\ and\ \citenamefont
  {Yoshida}(2020)}]{Kasuya2020}%
  \BibitemOpen
  \bibfield  {author} {\bibinfo {author} {\bibfnamefont {H.}~\bibnamefont
  {Kasuya}}\ and\ \bibinfo {author} {\bibfnamefont {K.}~\bibnamefont
  {Yoshida}},\ }\href {https://doi.org/10.1093/ptep/ptaa163} {\bibfield
  {journal} {\bibinfo  {journal} {Progress of Theoretical and Experimental
  Physics}\ }\textbf {\bibinfo {volume} {2021}},\ \bibinfo {pages} {013D01}
  (\bibinfo {year} {2020})},\ \Eprint
  {https://arxiv.org/abs/https://academic.oup.com/ptep/article-pdf/2021/1/013D01/36123331/ptaa163.pdf}
  {https://academic.oup.com/ptep/article-pdf/2021/1/013D01/36123331/ptaa163.pdf}
  \BibitemShut {NoStop}%
\bibitem [{\citenamefont {Robledo}\ \emph {et~al.}(2019)\citenamefont
  {Robledo}, \citenamefont {Rodríguez},\ and\ \citenamefont
  {Rodríguez-Guzmán}}]{Robledo2019}%
  \BibitemOpen
  \bibfield  {author} {\bibinfo {author} {\bibfnamefont {L.~M.}\ \bibnamefont
  {Robledo}}, \bibinfo {author} {\bibfnamefont {T.~R.}\ \bibnamefont
  {Rodríguez}},\ and\ \bibinfo {author} {\bibfnamefont {R.~R.}\ \bibnamefont
  {Rodríguez-Guzmán}},\ }\href
  {http://stacks.iop.org/0954-3899/46/i=1/a=013001} {\bibfield  {journal}
  {\bibinfo  {journal} {Journal of Physics G: Nuclear and Particle Physics}\
  }\textbf {\bibinfo {volume} {46}},\ \bibinfo {pages} {013001} (\bibinfo
  {year} {2019})}\BibitemShut {NoStop}%
\bibitem [{\citenamefont {Tandel}\ \emph {et~al.}(2006)\citenamefont {Tandel},
  \citenamefont {Chowdhury}, \citenamefont {Seabury}, \citenamefont {Ahmad},
  \citenamefont {Carpenter}, \citenamefont {Fischer}, \citenamefont {Janssens},
  \citenamefont {Khoo}, \citenamefont {Lauritsen}, \citenamefont {Lister},
  \citenamefont {Seweryniak},\ and\ \citenamefont {Shimizu}}]{Tandel2006}%
  \BibitemOpen
  \bibfield  {author} {\bibinfo {author} {\bibfnamefont {S.~K.}\ \bibnamefont
  {Tandel}}, \bibinfo {author} {\bibfnamefont {P.}~\bibnamefont {Chowdhury}},
  \bibinfo {author} {\bibfnamefont {E.~H.}\ \bibnamefont {Seabury}}, \bibinfo
  {author} {\bibfnamefont {I.}~\bibnamefont {Ahmad}}, \bibinfo {author}
  {\bibfnamefont {M.~P.}\ \bibnamefont {Carpenter}}, \bibinfo {author}
  {\bibfnamefont {S.~M.}\ \bibnamefont {Fischer}}, \bibinfo {author}
  {\bibfnamefont {R.~V.~F.}\ \bibnamefont {Janssens}}, \bibinfo {author}
  {\bibfnamefont {T.~L.}\ \bibnamefont {Khoo}}, \bibinfo {author}
  {\bibfnamefont {T.}~\bibnamefont {Lauritsen}}, \bibinfo {author}
  {\bibfnamefont {C.~J.}\ \bibnamefont {Lister}}, \bibinfo {author}
  {\bibfnamefont {D.}~\bibnamefont {Seweryniak}},\ and\ \bibinfo {author}
  {\bibfnamefont {Y.~R.}\ \bibnamefont {Shimizu}},\ }\href
  {https://doi.org/10.1103/PhysRevC.73.044306} {\bibfield  {journal} {\bibinfo
  {journal} {Phys. Rev. C}\ }\textbf {\bibinfo {volume} {73}},\ \bibinfo
  {pages} {044306} (\bibinfo {year} {2006})}\BibitemShut {NoStop}%
\bibitem [{\citenamefont {Clark}\ \emph {et~al.}(2010)\citenamefont {Clark},
  \citenamefont {Gregorich}, \citenamefont {Berryman}, \citenamefont {Ali},
  \citenamefont {Allmond}, \citenamefont {Beausang}, \citenamefont {Cromaz},
  \citenamefont {Deleplanque}, \citenamefont {Dragojevi\"A}, \citenamefont
  {Dvorak}, \citenamefont {Ellison}, \citenamefont {Fallon}, \citenamefont
  {Garcia}, \citenamefont {Gates}, \citenamefont {Gros}, \citenamefont
  {Jeppesen}, \citenamefont {Kaji}, \citenamefont {Lee}, \citenamefont
  {Macchiavelli}, \citenamefont {Morimoto}, \citenamefont {Nitsche},
  \citenamefont {Paschalis}, \citenamefont {Petri}, \citenamefont {Stavsetra},
  \citenamefont {Stephens}, \citenamefont {Watanabe},\ and\ \citenamefont
  {Wiedeking}}]{Clark201019}%
  \BibitemOpen
  \bibfield  {author} {\bibinfo {author} {\bibfnamefont {R.}~\bibnamefont
  {Clark}}, \bibinfo {author} {\bibfnamefont {K.}~\bibnamefont {Gregorich}},
  \bibinfo {author} {\bibfnamefont {J.}~\bibnamefont {Berryman}}, \bibinfo
  {author} {\bibfnamefont {M.}~\bibnamefont {Ali}}, \bibinfo {author}
  {\bibfnamefont {J.}~\bibnamefont {Allmond}}, \bibinfo {author} {\bibfnamefont
  {C.}~\bibnamefont {Beausang}}, \bibinfo {author} {\bibfnamefont
  {M.}~\bibnamefont {Cromaz}}, \bibinfo {author} {\bibfnamefont
  {M.}~\bibnamefont {Deleplanque}}, \bibinfo {author} {\bibfnamefont
  {I.}~\bibnamefont {Dragojevi\"A}}, \bibinfo {author} {\bibfnamefont
  {J.}~\bibnamefont {Dvorak}}, \bibinfo {author} {\bibfnamefont
  {P.}~\bibnamefont {Ellison}}, \bibinfo {author} {\bibfnamefont
  {P.}~\bibnamefont {Fallon}}, \bibinfo {author} {\bibfnamefont
  {M.}~\bibnamefont {Garcia}}, \bibinfo {author} {\bibfnamefont
  {J.}~\bibnamefont {Gates}}, \bibinfo {author} {\bibfnamefont
  {S.}~\bibnamefont {Gros}}, \bibinfo {author} {\bibfnamefont {H.}~\bibnamefont
  {Jeppesen}}, \bibinfo {author} {\bibfnamefont {D.}~\bibnamefont {Kaji}},
  \bibinfo {author} {\bibfnamefont {I.}~\bibnamefont {Lee}}, \bibinfo {author}
  {\bibfnamefont {A.}~\bibnamefont {Macchiavelli}}, \bibinfo {author}
  {\bibfnamefont {K.}~\bibnamefont {Morimoto}}, \bibinfo {author}
  {\bibfnamefont {H.}~\bibnamefont {Nitsche}}, \bibinfo {author} {\bibfnamefont
  {S.}~\bibnamefont {Paschalis}}, \bibinfo {author} {\bibfnamefont
  {M.}~\bibnamefont {Petri}}, \bibinfo {author} {\bibfnamefont
  {L.}~\bibnamefont {Stavsetra}}, \bibinfo {author} {\bibfnamefont
  {F.}~\bibnamefont {Stephens}}, \bibinfo {author} {\bibfnamefont
  {H.}~\bibnamefont {Watanabe}},\ and\ \bibinfo {author} {\bibfnamefont
  {M.}~\bibnamefont {Wiedeking}},\ }\href
  {https://doi.org/http://dx.doi.org/10.1016/j.physletb.2010.04.079} {\bibfield
   {journal} {\bibinfo  {journal} {Physics Letters B}\ }\textbf {\bibinfo
  {volume} {690}},\ \bibinfo {pages} {19 } (\bibinfo {year}
  {2010})}\BibitemShut {NoStop}%
\bibitem [{\citenamefont {Berger}\ \emph {et~al.}(1984)\citenamefont {Berger},
  \citenamefont {Girod},\ and\ \citenamefont {Gogny}}]{Berger1984}%
  \BibitemOpen
  \bibfield  {author} {\bibinfo {author} {\bibfnamefont {J.~F.}\ \bibnamefont
  {Berger}}, \bibinfo {author} {\bibfnamefont {M.}~\bibnamefont {Girod}},\ and\
  \bibinfo {author} {\bibfnamefont {D.}~\bibnamefont {Gogny}},\ }\href@noop {}
  {\bibfield  {journal} {\bibinfo  {journal} {Nucl. Phys. A}\ }\textbf
  {\bibinfo {volume} {428}},\ \bibinfo {pages} {23} (\bibinfo {year}
  {1984})}\BibitemShut {NoStop}%
\bibitem [{\citenamefont {Goriely}\ \emph {et~al.}(2009)\citenamefont
  {Goriely}, \citenamefont {Hilaire}, \citenamefont {Girod},\ and\
  \citenamefont {P\'{e}ru}}]{Goriely2009}%
  \BibitemOpen
  \bibfield  {author} {\bibinfo {author} {\bibfnamefont {S.}~\bibnamefont
  {Goriely}}, \bibinfo {author} {\bibfnamefont {S.}~\bibnamefont {Hilaire}},
  \bibinfo {author} {\bibfnamefont {M.}~\bibnamefont {Girod}},\ and\ \bibinfo
  {author} {\bibfnamefont {S.}~\bibnamefont {P\'{e}ru}},\ }\href@noop {}
  {\bibfield  {journal} {\bibinfo  {journal} {Phys. Rev. Lett.}\ }\textbf
  {\bibinfo {volume} {102}},\ \bibinfo {pages} {242501} (\bibinfo {year}
  {2009})}\BibitemShut {NoStop}%
\bibitem [{\citenamefont {Gonzalez-Boquera}\ \emph {et~al.}(2018)\citenamefont
  {Gonzalez-Boquera}, \citenamefont {Centelles}, \citenamefont {Viñas},\ and\
  \citenamefont {Robledo}}]{GonzalezBoquera2018}%
  \BibitemOpen
  \bibfield  {author} {\bibinfo {author} {\bibfnamefont {C.}~\bibnamefont
  {Gonzalez-Boquera}}, \bibinfo {author} {\bibfnamefont {M.}~\bibnamefont
  {Centelles}}, \bibinfo {author} {\bibfnamefont {X.}~\bibnamefont {Viñas}},\
  and\ \bibinfo {author} {\bibfnamefont {L.}~\bibnamefont {Robledo}},\ }\href
  {https://doi.org/https://doi.org/10.1016/j.physletb.2018.02.005} {\bibfield
  {journal} {\bibinfo  {journal} {Physics Letters B}\ }\textbf {\bibinfo
  {volume} {779}},\ \bibinfo {pages} {195 } (\bibinfo {year}
  {2018})}\BibitemShut {NoStop}%
\bibitem [{\citenamefont {Chappert}\ \emph {et~al.}(2015)\citenamefont
  {Chappert}, \citenamefont {Pillet}, \citenamefont {Girod},\ and\
  \citenamefont {Berger}}]{Chappert2015}%
  \BibitemOpen
  \bibfield  {author} {\bibinfo {author} {\bibfnamefont {F.}~\bibnamefont
  {Chappert}}, \bibinfo {author} {\bibfnamefont {N.}~\bibnamefont {Pillet}},
  \bibinfo {author} {\bibfnamefont {M.}~\bibnamefont {Girod}},\ and\ \bibinfo
  {author} {\bibfnamefont {J.-F.}\ \bibnamefont {Berger}},\ }\href
  {https://doi.org/10.1103/PhysRevC.91.034312} {\bibfield  {journal} {\bibinfo
  {journal} {Phys. Rev. C}\ }\textbf {\bibinfo {volume} {91}},\ \bibinfo
  {pages} {034312} (\bibinfo {year} {2015})}\BibitemShut {NoStop}%
\bibitem [{\citenamefont {Batail}\ \emph {et~al.}(2023)\citenamefont {Batail},
  \citenamefont {Davesne}, \citenamefont {P{\'e}ru}, \citenamefont {Becker},
  \citenamefont {Pastore},\ and\ \citenamefont {Navarro}}]{Batail2023}%
  \BibitemOpen
  \bibfield  {author} {\bibinfo {author} {\bibfnamefont {L.}~\bibnamefont
  {Batail}}, \bibinfo {author} {\bibfnamefont {D.}~\bibnamefont {Davesne}},
  \bibinfo {author} {\bibfnamefont {S.}~\bibnamefont {P{\'e}ru}}, \bibinfo
  {author} {\bibfnamefont {P.}~\bibnamefont {Becker}}, \bibinfo {author}
  {\bibfnamefont {A.}~\bibnamefont {Pastore}},\ and\ \bibinfo {author}
  {\bibfnamefont {J.}~\bibnamefont {Navarro}},\ }\href
  {https://doi.org/10.1140/epja/s10050-023-01073-w} {\bibfield  {journal}
  {\bibinfo  {journal} {The European Physical Journal A}\ }\textbf {\bibinfo
  {volume} {59}},\ \bibinfo {pages} {173} (\bibinfo {year} {2023})}\BibitemShut
  {NoStop}%
\bibitem [{\citenamefont {Baldo}\ \emph {et~al.}(2023)\citenamefont {Baldo},
  \citenamefont {Robledo},\ and\ \citenamefont {Vi{\~{n}}as}}]{Baldo2023}%
  \BibitemOpen
  \bibfield  {author} {\bibinfo {author} {\bibfnamefont {M.}~\bibnamefont
  {Baldo}}, \bibinfo {author} {\bibfnamefont {L.~M.}\ \bibnamefont {Robledo}},\
  and\ \bibinfo {author} {\bibfnamefont {X.}~\bibnamefont {Vi{\~{n}}as}},\
  }\href {https://doi.org/10.1140/epja/s10050-023-01062-z} {\bibfield
  {journal} {\bibinfo  {journal} {Eur. Phys. J. A}\ }\textbf {\bibinfo {volume}
  {59}},\ \bibinfo {pages} {156} (\bibinfo {year} {2023})}\BibitemShut
  {NoStop}%
\bibitem [{\citenamefont {Perez-Martin}\ and\ \citenamefont
  {Robledo}(2008)}]{PerezMartin2008}%
  \BibitemOpen
  \bibfield  {author} {\bibinfo {author} {\bibfnamefont {S.}~\bibnamefont
  {Perez-Martin}}\ and\ \bibinfo {author} {\bibfnamefont {L.~M.}\ \bibnamefont
  {Robledo}},\ }\href {https://doi.org/10.1103/PhysRevC.78.014304} {\bibfield
  {journal} {\bibinfo  {journal} {Phys. Rev. C}\ }\textbf {\bibinfo {volume}
  {78}},\ \bibinfo {pages} {014304} (\bibinfo {year} {2008})}\BibitemShut
  {NoStop}%
\bibitem [{\citenamefont {Schunck}\ \emph {et~al.}(2010)\citenamefont
  {Schunck}, \citenamefont {Dobaczewski}, \citenamefont {McDonnell},
  \citenamefont {Mor\'e}, \citenamefont {Nazarewicz}, \citenamefont {Sarich},\
  and\ \citenamefont {Stoitsov}}]{Schunck2010}%
  \BibitemOpen
  \bibfield  {author} {\bibinfo {author} {\bibfnamefont {N.}~\bibnamefont
  {Schunck}}, \bibinfo {author} {\bibfnamefont {J.}~\bibnamefont
  {Dobaczewski}}, \bibinfo {author} {\bibfnamefont {J.}~\bibnamefont
  {McDonnell}}, \bibinfo {author} {\bibfnamefont {J.}~\bibnamefont {Mor\'e}},
  \bibinfo {author} {\bibfnamefont {W.}~\bibnamefont {Nazarewicz}}, \bibinfo
  {author} {\bibfnamefont {J.}~\bibnamefont {Sarich}},\ and\ \bibinfo {author}
  {\bibfnamefont {M.~V.}\ \bibnamefont {Stoitsov}},\ }\href
  {https://doi.org/10.1103/PhysRevC.81.024316} {\bibfield  {journal} {\bibinfo
  {journal} {Phys. Rev. C}\ }\textbf {\bibinfo {volume} {81}},\ \bibinfo
  {pages} {024316} (\bibinfo {year} {2010})}\BibitemShut {NoStop}%
\bibitem [{\citenamefont {Giuliani}\ and\ \citenamefont
  {Robledo}(2023)}]{Giuliani2023}%
  \BibitemOpen
  \bibfield  {author} {\bibinfo {author} {\bibfnamefont {S.}~\bibnamefont
  {Giuliani}}\ and\ \bibinfo {author} {\bibfnamefont {L.}~\bibnamefont
  {Robledo}},\ }\href@noop {} {\bibfield  {journal} {\bibinfo  {journal}
  {Submitted to Physical Review C}\ } (\bibinfo {year} {2023})}\BibitemShut
  {NoStop}%
\bibitem [{\citenamefont {Bohr}\ and\ \citenamefont
  {Mottelson}(1975)}]{Bohr1975}%
  \BibitemOpen
  \bibfield  {author} {\bibinfo {author} {\bibfnamefont {A.}~\bibnamefont
  {Bohr}}\ and\ \bibinfo {author} {\bibfnamefont {B.}~\bibnamefont
  {Mottelson}},\ }\href@noop {} {\emph {\bibinfo {title} {Nuclear
  Structure}}},\ Vol.~\bibinfo {volume} {II}\ (\bibinfo  {publisher}
  {Benjamin},\ \bibinfo {address} {New-York},\ \bibinfo {year}
  {1975})\BibitemShut {NoStop}%
\bibitem [{\citenamefont {Robledo}\ and\ \citenamefont
  {Bertsch}(2012)}]{Robledo2012}%
  \BibitemOpen
  \bibfield  {author} {\bibinfo {author} {\bibfnamefont {L.~M.}\ \bibnamefont
  {Robledo}}\ and\ \bibinfo {author} {\bibfnamefont {G.~F.}\ \bibnamefont
  {Bertsch}},\ }\href {https://doi.org/10.1103/PhysRevC.86.054306} {\bibfield
  {journal} {\bibinfo  {journal} {Phys. Rev. C}\ }\textbf {\bibinfo {volume}
  {86}},\ \bibinfo {pages} {054306} (\bibinfo {year} {2012})}\BibitemShut
  {NoStop}%
\bibitem [{\citenamefont {Sheikh}\ \emph {et~al.}(2021)\citenamefont {Sheikh},
  \citenamefont {Dobaczewski}, \citenamefont {Ring}, \citenamefont {Robledo},\
  and\ \citenamefont {Yannouleas}}]{Sheikh2021}%
  \BibitemOpen
  \bibfield  {author} {\bibinfo {author} {\bibfnamefont {J.~A.}\ \bibnamefont
  {Sheikh}}, \bibinfo {author} {\bibfnamefont {J.}~\bibnamefont {Dobaczewski}},
  \bibinfo {author} {\bibfnamefont {P.}~\bibnamefont {Ring}}, \bibinfo {author}
  {\bibfnamefont {L.~M.}\ \bibnamefont {Robledo}},\ and\ \bibinfo {author}
  {\bibfnamefont {C.}~\bibnamefont {Yannouleas}},\ }\href
  {https://doi.org/10.1088/1361-6471/ac288a} {\bibfield  {journal} {\bibinfo
  {journal} {Journal of Physics G: Nuclear and Particle Physics}\ }\textbf
  {\bibinfo {volume} {48}},\ \bibinfo {pages} {123001} (\bibinfo {year}
  {2021})}\BibitemShut {NoStop}%
\bibitem [{\citenamefont {Dobaczewski}\ \emph {et~al.}(2015)\citenamefont
  {Dobaczewski}, \citenamefont {Afanasjev}, \citenamefont {Bender},
  \citenamefont {Robledo},\ and\ \citenamefont {Shi}}]{Dobaczewski2015}%
  \BibitemOpen
  \bibfield  {author} {\bibinfo {author} {\bibfnamefont {J.}~\bibnamefont
  {Dobaczewski}}, \bibinfo {author} {\bibfnamefont {A.}~\bibnamefont
  {Afanasjev}}, \bibinfo {author} {\bibfnamefont {M.}~\bibnamefont {Bender}},
  \bibinfo {author} {\bibfnamefont {L.}~\bibnamefont {Robledo}},\ and\ \bibinfo
  {author} {\bibfnamefont {Y.}~\bibnamefont {Shi}},\ }\href
  {https://doi.org/https://doi.org/10.1016/j.nuclphysa.2015.07.015} {\bibfield
  {journal} {\bibinfo  {journal} {Nuclear Physics A}\ }\textbf {\bibinfo
  {volume} {944}},\ \bibinfo {pages} {388} (\bibinfo {year} {2015})},\ \bibinfo
  {note} {special Issue on Superheavy Elements}\BibitemShut {NoStop}%
\bibitem [{\citenamefont {Almehed}\ \emph {et~al.}(2001)\citenamefont
  {Almehed}, \citenamefont {Frauendorf},\ and\ \citenamefont
  {D\"onau}}]{Almehed2001}%
  \BibitemOpen
  \bibfield  {author} {\bibinfo {author} {\bibfnamefont {D.}~\bibnamefont
  {Almehed}}, \bibinfo {author} {\bibfnamefont {S.}~\bibnamefont
  {Frauendorf}},\ and\ \bibinfo {author} {\bibfnamefont {F.}~\bibnamefont
  {D\"onau}},\ }\href {https://doi.org/10.1103/PhysRevC.63.044311} {\bibfield
  {journal} {\bibinfo  {journal} {Phys. Rev. C}\ }\textbf {\bibinfo {volume}
  {63}},\ \bibinfo {pages} {044311} (\bibinfo {year} {2001})}\BibitemShut
  {NoStop}%
\bibitem [{\citenamefont {Cullen}\ \emph {et~al.}(1999)\citenamefont {Cullen},
  \citenamefont {King}, \citenamefont {Reed}, \citenamefont {Sampson},
  \citenamefont {Walker}, \citenamefont {Wheldon}, \citenamefont {Xu},
  \citenamefont {Dracoulis}, \citenamefont {Lee}, \citenamefont {Macchiavelli},
  \citenamefont {MacLeod}, \citenamefont {Wilson},\ and\ \citenamefont
  {Barton}}]{Cullen1999}%
  \BibitemOpen
  \bibfield  {author} {\bibinfo {author} {\bibfnamefont {D.~M.}\ \bibnamefont
  {Cullen}}, \bibinfo {author} {\bibfnamefont {S.~L.}\ \bibnamefont {King}},
  \bibinfo {author} {\bibfnamefont {A.~T.}\ \bibnamefont {Reed}}, \bibinfo
  {author} {\bibfnamefont {J.~A.}\ \bibnamefont {Sampson}}, \bibinfo {author}
  {\bibfnamefont {P.~M.}\ \bibnamefont {Walker}}, \bibinfo {author}
  {\bibfnamefont {C.}~\bibnamefont {Wheldon}}, \bibinfo {author} {\bibfnamefont
  {F.}~\bibnamefont {Xu}}, \bibinfo {author} {\bibfnamefont {G.~D.}\
  \bibnamefont {Dracoulis}}, \bibinfo {author} {\bibfnamefont {I.-Y.}\
  \bibnamefont {Lee}}, \bibinfo {author} {\bibfnamefont {A.~O.}\ \bibnamefont
  {Macchiavelli}}, \bibinfo {author} {\bibfnamefont {R.~W.}\ \bibnamefont
  {MacLeod}}, \bibinfo {author} {\bibfnamefont {A.~N.}\ \bibnamefont
  {Wilson}},\ and\ \bibinfo {author} {\bibfnamefont {C.}~\bibnamefont
  {Barton}},\ }\href {https://doi.org/10.1103/PhysRevC.60.064301} {\bibfield
  {journal} {\bibinfo  {journal} {Phys. Rev. C}\ }\textbf {\bibinfo {volume}
  {60}},\ \bibinfo {pages} {064301} (\bibinfo {year} {1999})}\BibitemShut
  {NoStop}%
\bibitem [{\citenamefont {Jiao}\ \emph {et~al.}(2012)\citenamefont {Jiao},
  \citenamefont {Shi}, \citenamefont {Xu}, \citenamefont {Sun},\ and\
  \citenamefont {Walker}}]{Jiao2012}%
  \BibitemOpen
  \bibfield  {author} {\bibinfo {author} {\bibfnamefont {C.}~\bibnamefont
  {Jiao}}, \bibinfo {author} {\bibfnamefont {Y.}~\bibnamefont {Shi}}, \bibinfo
  {author} {\bibfnamefont {F.}~\bibnamefont {Xu}}, \bibinfo {author}
  {\bibfnamefont {Y.}~\bibnamefont {Sun}},\ and\ \bibinfo {author}
  {\bibfnamefont {P.~M.}\ \bibnamefont {Walker}},\ }\href
  {https://doi.org/10.1007/s11433-012-4824-4} {\bibfield  {journal} {\bibinfo
  {journal} {Science China Physics, Mechanics and Astronomy}\ }\textbf
  {\bibinfo {volume} {55}},\ \bibinfo {pages} {1613} (\bibinfo {year}
  {2012})}\BibitemShut {NoStop}%
\bibitem [{\citenamefont {Regan}\ \emph {et~al.}(1994)\citenamefont {Regan},
  \citenamefont {Walker}, \citenamefont {Dracoulis}, \citenamefont {Anderssen},
  \citenamefont {Byrne}, \citenamefont {Davidson}, \citenamefont {Kibèdi},
  \citenamefont {Lane}, \citenamefont {Stuchbery},\ and\ \citenamefont
  {Yeung}}]{Regan1994}%
  \BibitemOpen
  \bibfield  {author} {\bibinfo {author} {\bibfnamefont {P.}~\bibnamefont
  {Regan}}, \bibinfo {author} {\bibfnamefont {P.}~\bibnamefont {Walker}},
  \bibinfo {author} {\bibfnamefont {G.}~\bibnamefont {Dracoulis}}, \bibinfo
  {author} {\bibfnamefont {S.}~\bibnamefont {Anderssen}}, \bibinfo {author}
  {\bibfnamefont {A.}~\bibnamefont {Byrne}}, \bibinfo {author} {\bibfnamefont
  {P.}~\bibnamefont {Davidson}}, \bibinfo {author} {\bibfnamefont
  {T.}~\bibnamefont {Kibèdi}}, \bibinfo {author} {\bibfnamefont
  {G.}~\bibnamefont {Lane}}, \bibinfo {author} {\bibfnamefont {A.}~\bibnamefont
  {Stuchbery}},\ and\ \bibinfo {author} {\bibfnamefont {K.}~\bibnamefont
  {Yeung}},\ }\href
  {https://doi.org/https://doi.org/10.1016/0375-9474(94)90156-2} {\bibfield
  {journal} {\bibinfo  {journal} {Nuclear Physics A}\ }\textbf {\bibinfo
  {volume} {567}},\ \bibinfo {pages} {414} (\bibinfo {year}
  {1994})}\BibitemShut {NoStop}%
\bibitem [{\citenamefont {Shizuma}\ \emph {et~al.}(1995)\citenamefont
  {Shizuma}, \citenamefont {Mitarai}, \citenamefont {Sletten}, \citenamefont
  {Bark}, \citenamefont {Gjørup}, \citenamefont {Jensen}, \citenamefont
  {Wrzesinski},\ and\ \citenamefont {Piiparinen}}]{Shizuma1995}%
  \BibitemOpen
  \bibfield  {author} {\bibinfo {author} {\bibfnamefont {T.}~\bibnamefont
  {Shizuma}}, \bibinfo {author} {\bibfnamefont {S.}~\bibnamefont {Mitarai}},
  \bibinfo {author} {\bibfnamefont {G.}~\bibnamefont {Sletten}}, \bibinfo
  {author} {\bibfnamefont {R.}~\bibnamefont {Bark}}, \bibinfo {author}
  {\bibfnamefont {N.}~\bibnamefont {Gjørup}}, \bibinfo {author} {\bibfnamefont
  {H.}~\bibnamefont {Jensen}}, \bibinfo {author} {\bibfnamefont
  {J.}~\bibnamefont {Wrzesinski}},\ and\ \bibinfo {author} {\bibfnamefont
  {M.}~\bibnamefont {Piiparinen}},\ }\href
  {https://doi.org/https://doi.org/10.1016/0375-9474(95)00376-C} {\bibfield
  {journal} {\bibinfo  {journal} {Nuclear Physics A}\ }\textbf {\bibinfo
  {volume} {593}},\ \bibinfo {pages} {247} (\bibinfo {year}
  {1995})}\BibitemShut {NoStop}%
\bibitem [{\citenamefont {Purry}\ \emph {et~al.}(1998)\citenamefont {Purry},
  \citenamefont {Walker}, \citenamefont {Dracoulis}, \citenamefont {Kibédi},
  \citenamefont {Kondev}, \citenamefont {Bayer}, \citenamefont {Bruce},
  \citenamefont {Byrne}, \citenamefont {Gelletly}, \citenamefont {Regan},
  \citenamefont {Thwaites}, \citenamefont {Burglin},\ and\ \citenamefont
  {Rowley}}]{Purry1998}%
  \BibitemOpen
  \bibfield  {author} {\bibinfo {author} {\bibfnamefont {C.}~\bibnamefont
  {Purry}}, \bibinfo {author} {\bibfnamefont {P.}~\bibnamefont {Walker}},
  \bibinfo {author} {\bibfnamefont {G.}~\bibnamefont {Dracoulis}}, \bibinfo
  {author} {\bibfnamefont {T.}~\bibnamefont {Kibédi}}, \bibinfo {author}
  {\bibfnamefont {F.}~\bibnamefont {Kondev}}, \bibinfo {author} {\bibfnamefont
  {S.}~\bibnamefont {Bayer}}, \bibinfo {author} {\bibfnamefont
  {A.}~\bibnamefont {Bruce}}, \bibinfo {author} {\bibfnamefont
  {A.}~\bibnamefont {Byrne}}, \bibinfo {author} {\bibfnamefont
  {W.}~\bibnamefont {Gelletly}}, \bibinfo {author} {\bibfnamefont
  {P.}~\bibnamefont {Regan}}, \bibinfo {author} {\bibfnamefont
  {C.}~\bibnamefont {Thwaites}}, \bibinfo {author} {\bibfnamefont
  {O.}~\bibnamefont {Burglin}},\ and\ \bibinfo {author} {\bibfnamefont
  {N.}~\bibnamefont {Rowley}},\ }\href
  {https://doi.org/https://doi.org/10.1016/S0375-9474(97)00654-4} {\bibfield
  {journal} {\bibinfo  {journal} {Nuclear Physics A}\ }\textbf {\bibinfo
  {volume} {632}},\ \bibinfo {pages} {229} (\bibinfo {year}
  {1998})}\BibitemShut {NoStop}%
\bibitem [{\citenamefont {Baglin}(2003)}]{Baglin2003}%
  \BibitemOpen
  \bibfield  {author} {\bibinfo {author} {\bibfnamefont {C.~M.}\ \bibnamefont
  {Baglin}},\ }\href {https://doi.org/https://doi.org/10.1006/ndsh.2003.0007}
  {\bibfield  {journal} {\bibinfo  {journal} {Nuclear Data Sheets}\ }\textbf
  {\bibinfo {volume} {99}},\ \bibinfo {pages} {1} (\bibinfo {year}
  {2003})}\BibitemShut {NoStop}%
\bibitem [{\citenamefont {Achterberg}\ \emph {et~al.}(2009)\citenamefont
  {Achterberg}, \citenamefont {Capurro},\ and\ \citenamefont
  {Marti}}]{Achterberg2009}%
  \BibitemOpen
  \bibfield  {author} {\bibinfo {author} {\bibfnamefont {E.}~\bibnamefont
  {Achterberg}}, \bibinfo {author} {\bibfnamefont {O.}~\bibnamefont
  {Capurro}},\ and\ \bibinfo {author} {\bibfnamefont {G.}~\bibnamefont
  {Marti}},\ }\href {https://doi.org/https://doi.org/10.1016/j.nds.2009.05.002}
  {\bibfield  {journal} {\bibinfo  {journal} {Nuclear Data Sheets}\ }\textbf
  {\bibinfo {volume} {110}},\ \bibinfo {pages} {1473} (\bibinfo {year}
  {2009})}\BibitemShut {NoStop}%
\bibitem [{\citenamefont {Singh}\ and\ \citenamefont
  {Roediger}(2010)}]{Singh2010}%
  \BibitemOpen
  \bibfield  {author} {\bibinfo {author} {\bibfnamefont {B.}~\bibnamefont
  {Singh}}\ and\ \bibinfo {author} {\bibfnamefont {J.~C.}\ \bibnamefont
  {Roediger}},\ }\href
  {https://doi.org/https://doi.org/10.1016/j.nds.2010.08.001} {\bibfield
  {journal} {\bibinfo  {journal} {Nuclear Data Sheets}\ }\textbf {\bibinfo
  {volume} {111}},\ \bibinfo {pages} {2081} (\bibinfo {year}
  {2010})}\BibitemShut {NoStop}%
\bibitem [{\citenamefont {Meyer}\ \emph {et~al.}(2006)\citenamefont {Meyer},
  \citenamefont {Wood}, \citenamefont {Casten}, \citenamefont {Fitzpatrick},
  \citenamefont {Graw}, \citenamefont {Bucurescu}, \citenamefont {Jolie},
  \citenamefont {{von Brentano}}, \citenamefont {Hertenberger}, \citenamefont
  {Wirth}, \citenamefont {Braun}, \citenamefont {Faestermann}, \citenamefont
  {Heinze}, \citenamefont {Jerke}, \citenamefont {Krücken}, \citenamefont
  {Mahgoub}, \citenamefont {Möller}, \citenamefont {Mücher},\ and\
  \citenamefont {Scholl}}]{Meyer2006}%
  \BibitemOpen
  \bibfield  {author} {\bibinfo {author} {\bibfnamefont {D.}~\bibnamefont
  {Meyer}}, \bibinfo {author} {\bibfnamefont {V.}~\bibnamefont {Wood}},
  \bibinfo {author} {\bibfnamefont {R.}~\bibnamefont {Casten}}, \bibinfo
  {author} {\bibfnamefont {C.}~\bibnamefont {Fitzpatrick}}, \bibinfo {author}
  {\bibfnamefont {G.}~\bibnamefont {Graw}}, \bibinfo {author} {\bibfnamefont
  {D.}~\bibnamefont {Bucurescu}}, \bibinfo {author} {\bibfnamefont
  {J.}~\bibnamefont {Jolie}}, \bibinfo {author} {\bibfnamefont
  {P.}~\bibnamefont {{von Brentano}}}, \bibinfo {author} {\bibfnamefont
  {R.}~\bibnamefont {Hertenberger}}, \bibinfo {author} {\bibfnamefont {H.-F.}\
  \bibnamefont {Wirth}}, \bibinfo {author} {\bibfnamefont {N.}~\bibnamefont
  {Braun}}, \bibinfo {author} {\bibfnamefont {T.}~\bibnamefont {Faestermann}},
  \bibinfo {author} {\bibfnamefont {S.}~\bibnamefont {Heinze}}, \bibinfo
  {author} {\bibfnamefont {J.}~\bibnamefont {Jerke}}, \bibinfo {author}
  {\bibfnamefont {R.}~\bibnamefont {Krücken}}, \bibinfo {author}
  {\bibfnamefont {M.}~\bibnamefont {Mahgoub}}, \bibinfo {author} {\bibfnamefont
  {O.}~\bibnamefont {Möller}}, \bibinfo {author} {\bibfnamefont
  {D.}~\bibnamefont {Mücher}},\ and\ \bibinfo {author} {\bibfnamefont
  {C.}~\bibnamefont {Scholl}},\ }\href
  {https://doi.org/https://doi.org/10.1016/j.physletb.2006.05.007} {\bibfield
  {journal} {\bibinfo  {journal} {Physics Letters B}\ }\textbf {\bibinfo
  {volume} {638}},\ \bibinfo {pages} {44} (\bibinfo {year} {2006})}\BibitemShut
  {NoStop}%
\bibitem [{\citenamefont {Hofmann}\ \emph {et~al.}(2001)\citenamefont
  {Hofmann}, \citenamefont {He{\ss}berger}, \citenamefont {Ackermann},
  \citenamefont {Antalic}, \citenamefont {Cagarda}, \citenamefont
  {{\'{C}}wiok}, \citenamefont {Kindler}, \citenamefont {Kojouharova},
  \citenamefont {Lommel}, \citenamefont {Mann}, \citenamefont {M{\"u}nzenberg},
  \citenamefont {Popeko}, \citenamefont {Saro}, \citenamefont {Sch{\"o}tt},\
  and\ \citenamefont {Yeremin}}]{Hofmann2001}%
  \BibitemOpen
  \bibfield  {author} {\bibinfo {author} {\bibfnamefont {S.}~\bibnamefont
  {Hofmann}}, \bibinfo {author} {\bibfnamefont {F.~P.}\ \bibnamefont
  {He{\ss}berger}}, \bibinfo {author} {\bibfnamefont {D.}~\bibnamefont
  {Ackermann}}, \bibinfo {author} {\bibfnamefont {S.}~\bibnamefont {Antalic}},
  \bibinfo {author} {\bibfnamefont {P.}~\bibnamefont {Cagarda}}, \bibinfo
  {author} {\bibfnamefont {S.}~\bibnamefont {{\'{C}}wiok}}, \bibinfo {author}
  {\bibfnamefont {B.}~\bibnamefont {Kindler}}, \bibinfo {author} {\bibfnamefont
  {J.}~\bibnamefont {Kojouharova}}, \bibinfo {author} {\bibfnamefont
  {B.}~\bibnamefont {Lommel}}, \bibinfo {author} {\bibfnamefont
  {R.}~\bibnamefont {Mann}}, \bibinfo {author} {\bibfnamefont {G.}~\bibnamefont
  {M{\"u}nzenberg}}, \bibinfo {author} {\bibfnamefont {A.~G.}\ \bibnamefont
  {Popeko}}, \bibinfo {author} {\bibfnamefont {S.}~\bibnamefont {Saro}},
  \bibinfo {author} {\bibfnamefont {H.~J.}\ \bibnamefont {Sch{\"o}tt}},\ and\
  \bibinfo {author} {\bibfnamefont {A.~V.}\ \bibnamefont {Yeremin}},\ }\href
  {https://doi.org/10.1007/s100500170137} {\bibfield  {journal} {\bibinfo
  {journal} {The European Physical Journal A - Hadrons and Nuclei}\ }\textbf
  {\bibinfo {volume} {10}},\ \bibinfo {pages} {5} (\bibinfo {year}
  {2001})}\BibitemShut {NoStop}%
\end{thebibliography}

\end{document}